\RequirePackage{fix-cm}
\documentclass[twocolumn,epjc3]{svjour3}  
\smartqed  
\pdfoutput=1 

\usepackage[T1]{fontenc} 
\usepackage{color}   
\usepackage{graphicx,float}
\usepackage{dcolumn}
\usepackage{bm}
\usepackage{slashed}
\usepackage{float}
\usepackage{amsmath}
\usepackage{amssymb}

\journalname{{}}
\begin{document}

\def\bb    #1{\hbox{\boldmath${#1}$}}
 \def\oo    #1{{#1}_0 \!\!\!\!\!{}^{{}^{\circ}}~}  
 \def\op    #1{{#1}_0 \!\!\!\!\!{}^{{}^{{}^{\circ}}}~}
\def\blambda{{\hbox{\boldmath $\lambda$}}} 
\def\eeta{{\hbox{\boldmath $\eta$}}}
\def\bxi{{\hbox{\boldmath $\xi$}}} 
\def\bzeta{{\hbox{\boldmath $\zeta$}}}
\def\sD{D \!\!\!\!/}
\def\sd{\partial \!\!\!\!/}
\def\qcdu{{{}_{ \rm QCD}}}   
\def\qedu{{{}_{\rm QED}}}   
\def\qcdd{{{}^{ \rm QCD}}}   
\def\qedd{{{}^{\rm QED}}}   
\def\qcd{{{\rm QCD}}}   
\def\qed{{{\rm QED}}}   
\def\2d{{{}_{\rm 2D}}}         
\def\4d{{{}_{\rm 4D}}}         

\title{\boldmath On the stability of the open-string QED neutron and
  dark matter}


\titlerunning{QED neutron}        

\author{Cheuk-Yin Wong\thanksref{e1}}

\thankstext{t1}{This manuscript has been authored in part by
  UT-Battelle, LLC, under contract DE-AC05-00OR22725 with the US
  Department of Energy (DOE). The US government retains and the
  publisher, by accepting the article for publication, acknowledges
  that the US government retains a nonexclusive, paid-up, irrevocable,
  worldwide license to publish or reproduce the published form of this
  manuscript, or allow others to do so, for US government
  purposes. DOE will provide public access to these results of
  federally sponsored research in accordance with the DOE Public
  Access Plan (http://energy.gov/downloads/doe-public-access-plan),
  Oak Ridge, Tennessee 37831, USA} \thankstext{e1}{e-mail:
  wongc@ornl.gov}


\institute{Physics Division, Oak Ridge National Laboratory$^a$,
           Oak Ridge, Tennessee 37831, USA 
}

\date{}

\maketitle

\begin{abstract}
{
We study the stability of a hypothetical QED neutron, which consists
of a color-singlet system of two $d$ quarks and a $u$ quark
interacting with the QED interaction.  As a quark cannot be isolated,
the intrinsic motion of the three quarks in the lowest-energy state
may lie predominantly in 1+1 dimensions, as in a \break $d$-$u$-$d$
open string.  The attractive $d$-$u$ and $u$-$d$ QED interactions may
overcome the weaker repulsive $d$-$d$ QED interaction to bind the
three quarks together.  We examine the QED neutron in a
phenomenological three-body problem in 1+1 dimensions with an
effective interaction extracted from Schwinger's exact QED solution in
1+1 dimensions.  The phenomenological model in a variational
calculation yields a stable QED neutron at 44.5 MeV.  The analogous
QED proton with two $u$ quarks and a $d$ quark has been found to be
too repulsive to be stable and does not have a bound or continuum
state, onto which the QED neutron can decay via the weak interaction.
Consequently, the QED neutron is stable against the weak decay, has a
long lifetime, and is in fact a QED dark neutron. It may be produced
following the deconfinement-to-confinement phase transition of the
quark gluon plasma in high-energy heavy-ion collisions.  Because of
the long lifetime of the QED dark neutron, self-gravitating assemblies
of QED dark neutrons or dark antineutrons may be good candidates for a
part of the primordial dark matter produced during the phase
transition of the quark gluon plasma in the evolution of the early
Universe.  }

\keywords{Anomalous soft photons \and X17 \and E38 \and Schwinger QED2
  \and Open string \and Dark matter}

\end{abstract}

\maketitle \flushbottom

\setcounter{footnote}{0}

\section{Introduction}

Recent experimental observations of (i) the anomalous soft photons
\cite{Chl84,Bot91,Ban93,Bel97,Bel02,DEL06,DEL08,Per09,DEL10}, (ii) the
X17 particle at about 17 MeV \cite{Kra16,Kra19,Jai07}, and (iii) the
E38 particle at about 38 MeV \cite{Abr12,Abr19,Bev11,Bev12} have generated a
great deal of interests
\cite{Van89}-\cite{Dya21}.  With a mass in
the region of many tens of MeV, the produced anomalous particles
appear to lie outside the domain of the Standard Model.  Many
speculations have been proposed for these objects, including the cold
quark-gluon plasma, QED mesons, the fifth force of Nature, the
extension of the Standard Model, the QCD axion, dark matter and many
others \cite{X1722}.

Among the many proposed explanations for the above three anomalies,
the description of the quantized QED mesons
\cite{Won10,Won11,Won14,Won20,Won22a,Won22b,Kos21} has the prospect of
linking them together in a consistent framework.  We note that the
anomalous soft photons are consistently produced as observed excess
$e^+e^-$ pairs when hadrons are produced, and they are not produced
when hadrons are not produced
\cite{Chl84,Bot91,Ban93,Bel97,Bel02,DEL06,DEL08,Per09,DEL10}.  The
correlated production of anomalous soft photons alongside with hadrons
suggests that a parent particle of the anomalous soft photons is
likely to contain some elements of the hadrons, such as a light
quark-antiquark pair.  The transverse momentum range of the anomalous
soft photons of many tens of MeV/c suggests further that the parent
particle is likely the QED excitation of such a quark-antiquark pair.
For, it was shown by Schwinger previously that massless fermions (and
their antiparticles) interacting in a gauge field with a coupling
constant $g_\2d$ in 1+1 dimensions lead to stable bosons with a mass
$m=g_\2d/\sqrt{\pi}$ \cite{Sch62,Sch63}.  A Schwinger boson can be
viewed as the stable collective excitation of the fermions and their
gauge fields in the vacuum.  It can be alternatively viewed as a
confined open-string state of a fermion-antifermion pair bound
together by their mutual gauge field interaction.  We shall review
Schwinger's boson in (1+1)D QED in Appendix A and apply it to quarks
interacting in QED and QCD in Appendix B.  Thus, if we treat the light
quarks as Schwinger's massless fermions subject to QED and QCD gauge
interactions as discussed in Appendix A and B, the ratio of the mass
of a $q\bar q$ composite of the QED excitation (a QED meson) to the
mass of a $q\bar q$ composite of the QCD excitation (a QCD meson) as
given by Eq.\ (\ref{b16}) of Appendix B would be of order
\begin{eqnarray}
\frac{m_{\qedu\,{\rm meson}}}{m_{\qcdu\,{\rm
    meson}} }
    \sim \frac{g^\qedu}{g^\qcdu} 
    \sim \!\sqrt{\frac{\alpha_{{}_\qedd}}{\alpha_{{}_\qcdd}}}
    \sim \!\sqrt{\frac{1/137 }{0.6 }}
     \sim \frac{1}{9},
\end{eqnarray}
placing the QED meson within the domain of the anomalous soft photons.
It was therefore proposed in \cite{Won10,Won11,Won14,Won20} that QED
excitations of quarks may lead to confined, open-string QED mesons
with masses of many tens of MeV.  These QED mesons may be produced
simultaneously along with the QCD mesons when the quark system is
excited in high-energy collisions
\cite{Chl84,Bot91,Ban93,Bel97,Bel02,DEL06,DEL08,Per09,DEL10}, and the
excess $e^+e^-$ pairs may arise from the decays of these QED mesons.
By examining both the QCD string and the QED string excitations, using
the method of bosonization
\cite{Col75,Col76,Hal75,Cas74,Wit84,Fri93,Won95,Hos98,Abd01,Nag09} and
extrapolating from the QCD string excitations to the QED string
excitations, the mass of the $I(J^\pi)$=0(0$^-)$ isoscalar QED meson
was predicted to be 17.9$\pm$1.8 MeV and the mass of the isovector
$(I(J^\pi)$=1($0^-), I_3$=0) QED meson to be 36.4$\pm$3.8 MeV
\cite{Won20}.  These QED meson masses match those of the X17 particle,
the E38 particle, and the possible parent particles of the anomalous
soft photons, indicating that these anomalous particles are consistent
with their description as QED excitations of the quark vacuum.

It is commonly argued that quarks experience QCD and QED interactions
simultaneously and that the quark confinement is a property of quarks
in QCD alone.  It may appear at first sight that the above suggestion
of stable and QED excitations of confining quarks may appear
impossible.  Is it ever possible for quarks (and antiquarks) to
interact with the QED interaction alone without the QCD interaction?
Can there be stable collective QED excitations of quarks and
antiquarks, similar to the stable collective QCD quark-antiquark
excitations?  If quarks can interact with the QED interaction alone,
can there be additionally other similarly favorable neutral mutiquark
systems stabilized by the QED interaction?

With regard to the first question whether it is ever possible for
quarks (and antiquarks) to interact with the QED interaction alone, we
can consider as an example the production of a quark and an antiquark
pair by\break $e^+ + e^- \to \gamma^* \to q +\bar q$ ~or ~$e^+ + e^-
\to \gamma^* +\gamma^* \to q $+$\bar q$ \break with a center-of-mass
energy $\sqrt{s}$ in the range\break $(m_q + m_{\bar q}) <\sqrt{s} <
m_\pi$, where the sum of the rest masses of the light quark and light
antiquark is of order a few MeV and $m_\pi\sim 135$ MeV \cite{PDG19}.
The incident $e^+$+$e^-$ pair is in a colorless color-singlet state,
and thus the produced $q$-$\bar q$ pair is also a color-singlet pair.
The produced $q$-$\bar q$ pair, if they can ever be produced in this
range of $\sqrt{s}$, can only interact with the QED interaction but
not with the QCD interaction, because the QCD interaction would result
in a composite $q$-$\bar q$ hadron state with a center-of-mass energy
$\sqrt{s}$ beyond this energy range, in a contradictory manner.  It is
therefore possible for a quark and an antiquark to interact with QED
interaction alone, if they are produced in the range of $\sqrt{s}$
below the pion mass.  If the produced quark and the antiquark are not
confined by the QED interaction, the produced quark-antiquark pair
will appear as a continuum state with separated fractional charges of
a quark and an antiquark, at energies away from the bound state
energies.  The non-observation of fractional charges for the
$e^+$+$e^-$$\to$ $q$+$\bar q$ reaction in this $\sqrt{s}$ range away
from the bound state energies, is consistent with the confinement of
quarks in QED interactions in (3+1)D.
 
In other reactions, there are circumstances in which a $q$-$\bar q$
pair can be produced with a center-of-mass energy $\sqrt{s}$ lower
than the pion mass and can interact in the QED interaction alone
without the QCD interaction.  For example, we mentioned earlier the
production of a $q$-$\bar q$ pair interacting in QED interactions as a
possible description for the anomalous soft photons, the X17 particle,
and the E38 particle with masses in the range of many tens of MeV, as
described in \cite{Won20}.  In other circumstances in the
deconfinement-to-confinement phase transition of the qaurk-gluon
plasma in high-energy heavy-ion collisions, a deconfined quarks and a
deconfined antiquark in close spatial proximity can coalesce to become
a $q$-$\bar q$ pair with a pair energy below the pion mass, and they
can interact in QED interactions alone.

In the dynamics of quarks and antiquarks in a general reaction, the
quark currents and the gauge fields are not single-element functions.
They are in fact 3$\times 3$ color matrices.  Quarks in triplet $\bb
3$ representation and antiquarks in $\bb 3^*$ representation form the
product group of $\bb 3$ $\otimes$ $\bb 3^*$=$\bb 1$ $\oplus$ $\bb 8$,
which contains the color-singlet $\bb 1$ subgroup and the color-octet
$\bb 8$ subgroup.  Quark currents $j^\mu$ and gauge fields $A^\mu$ in
the color-singlet and color-octet subgroups execute collective
dynamics within their respective subgroups, as discussed in Appendix A
and B.  Quarks and antiquarks can therefore interact with the QED
interaction alone within the color-singlet subgroup, if their
center-of-mass energy is below the pion mass.

To describe the quark-QCD-QED dynamics in detail, we envisage the
vacuum to comprise of quarks filling up the negative-energy Dirac sea,
with transient quark-antiquark pairs, gluons, and photons lurking here
and disappearing there in the space time arena \cite{Gee10,Bed20}.  On
top of this vacuum, we introduce initial gauge field disturbances
$A_\qedd^\mu(x)$ and ${ A}_\qcdd^\mu(x)$ over a space-time region $x$,
which creates valence quarks occupying states above the Dirac sea and
valence antiquarks as unoccupied hole states below the Dirac sea.  The
interaction Lagrangian is \cite{Won10,Won20,Won22a,Won22b,Kos21}
\begin{eqnarray}
{\cal L}^{\rm int}&&={\cal L}_\qedd^{\rm int}+
{\cal L}_\qcdd^{\rm int},
\\
{\cal L}_\qedd^{\rm int} &&= \sum_{f,a',a} 
g^{\qedu} Q_f^\qedu
\bar \psi_{a'}^{f}  \gamma_\mu   (  A_\qedd^\mu )_{a' a}  \psi_a^f,
\\
 {\cal L}_\qcdd^{\rm int}&&= \sum_{f,a',a} g^{\qcdu} Q_f^\qcdu \bar \psi_{a'}^{f}  \gamma_\mu    (  { A}_\qcdd^\mu )_{a' a}  \psi_a^f,
\end{eqnarray}
where $\mu=0,1,2,3$ are the space-time index, $a=1,2,3$ is the color
label, $f$ is the flavor label, $g^{\qcdu,\qedu}$ are the coupling
constants, and $Q_f^{\{\qedu,\qcdu\}}$ is the charge number of
$f$-flavor quark in QED and QCD respectively.  In particular, the QED
gauge field $A_\qedd^\mu(x)$ consists of the component $A_0^\mu(x)$
associated with the generator $t^0$ of the QED U(1) gauge group,
\vspace*{-0.1cm}
\begin{eqnarray}
A_\qedd^\mu(x) = A_0^\mu(x) ~t^0, ~~~~~~ {\rm where} ~~t^0=\frac{1}{\sqrt{6}}
\left ( \begin{matrix}
                1 & 0 & 0 \cr
                0 & 1 & 0 \cr
                0 & 0 & 1
        \end{matrix}   \right )\!.
 \vspace*{-0.3cm}
\end{eqnarray}
The QCD gauge field ${ A}_\qcdd^\mu(x)$ consists of eight
components ${ A}_{1,2,...8}^\mu(x)$ associated with the eight
Gell-Mann matrices generators, $t^{1,2,...,8}$ of the QCD color SU(3)
gauge group,
\begin{eqnarray}
{ A}_\qcdd^\mu(x) = \sum_{i=1,2,...8} A_i^\mu(x)~ t^i,
~~~~~ 2\,{\rm Tr}\{ t^a t^b\}=\delta^{ab}. 
\end{eqnarray}
The initial QED and QCD gauge field disturbances will act on the quark
field through the Dirac equation
\begin{eqnarray}
\gamma_\mu\left  [ i \partial ^\mu + \sum_{i=0}^8 \sum_{a} g_f^i (A_i^\mu(x) t^i )_{a' a}\right ] \psi^f_a(x)=0
\end{eqnarray}
where $g_f^0$=\,$g^\qedu Q^\qedu_f$=\,$eQ^\qedu_f$, $
g_f^{1,2,3,..8}$=\,$g^\qcdu Q^\qcdu_f$,\break $Q^\qedu_u$=2/3,
$Q^\qedu_d$=$-$1/3, and $Q^\qcdu_{\{u,d,s\}}$=1.  The initial QED and
QCD gauge field disturbances will induce a change of the state vector
$\psi_a^f(x)$ of all quark states and subsequently a change of the
quark currents $j^{\mu'}(x)$.  Just as the gauge fields, the induced
quark currents $j^{\mu'}(x)$ are likewise a 3$\times$3 matrices in
color space which can be similarly described in terms of 9 independent
generators,
\begin{eqnarray}
j^{\mu'}(x) = \sum_{i=0,1,2,...8} j_i^{\mu'} (x)~t^i. 
\end{eqnarray}
In particular, they can be divided into the QED quark current
$j_\qedd^{\mu'}(x)$=$j_0^{\mu'} (x)t^0$ where
\begin{eqnarray}
 j_0^{\mu'}(x)  = \sum_{f,a',a}   g^\qedu Q^\qedu_f \bar \psi_{a'}^{f}(x') \gamma^{\mu'} (t^0 )_{a'a} \psi_a^f  (x)\bigr |_{x' \to x} ,
\end{eqnarray}
and the QCD quark currents $ j_\qcdd^{\mu'}(x)$=$\sum_{i=1,2,..8}
j_i^{\mu'} (x)t^i $ where for $i=1,2,..8$,
\begin{eqnarray}
 \!j_i^{\mu'}\!(x)\!=\! \!\sum_{f,a',a}\!\!  g^\qcdu Q^\qcdu_f  \bar \psi_{a'}^f (x') \gamma^{\mu'} (t^i )_{a'a} \psi_a^f (x)\bigr |_{x' \to x}.
\end{eqnarray}
Through the Maxwell equations for QCD and QED, the color-singlet QED
current $j_\qedd^{\mu'}$ will generate a color-singlet QED gauge
fields $\tilde A_\qedd^{\mu''}$, while the color-octet QCD currents ${
  j_\qcdd^{\mu'}}$ will generate a color-octet QCD gauge fields ${
  \tilde A_\qcdd^{\mu''}}$.  Along the evolution loop $A^\mu \to
j^{\mu'} \to \tilde A^{\mu''}\!,$ the requirement that the generated
gauge fields $\tilde A^{\mu''}$ at the end of loop be the same as the
gauge field disturbance $A^\mu$ initially introduced provides the
condition for stable QED and QCD collective excitations of the quark
vacuum as described in Appendix B
\cite{Won10,Won20,Won22a,Won22b,Kos21}.
  
For quarks in QCD, the idealization of a flux tube in
(3+1)D$_{\{x^1,x^2,x^3,x^0\}}$ as a one-dimensional string in
(1+1)D$_{\{x^3,x^0\}}$ is well known since the works of the dual
resonance model \cite{Ven68}, the Nambu-Goto string model
\cite{Nam70,Got71}, the t'Hoof two-dimensional model
\cite{tho74,tH74a,tH74b}, the inside-outside-cascade model
\cite{Cas74}, the yo-yo string model \cite{Art74} and the Lund model
\cite{And83}.  QCD lattice gauge calculations exhibit explicitly
the structure of a flux tube in 3+1 dimensions
\cite{Hua88,Bal05,Cos17}.

For quarks in QED, whether quarks are confined in the U(1) QED gauge
interaction in (3+1)D has not been unequivocally demonstrated,
although there have been strong hints pointing to the confinement of
quarks in QED if they can interact in the QED interaction alone.  The
principal cause of uncertainty arise from the fact that the QED U(1)
gauge field admits two different versions of gauge theories with
different confinement properties.  As emphasized by Polyakov
\cite{Pol77,Pol87} and Drell $et~al.$ \cite{Dre79}, there is the
compact U(1) QED gauge theory containing the gauge fields $A^\mu$ as
angular variables with periodic properties.  Defined on a lattice, the
compact U(1) QED gauge theory has the gauge field action
\cite{Pol77,Pol87}
\begin{eqnarray}
S=\frac{1}{2g^2}\sum_{x,\alpha \beta} (1-\cos F_{x,\alpha \beta}),
\end{eqnarray}
where $g$ is the coupling constant and $F_{x,\alpha \beta}$ is
\begin{subequations}
\begin{eqnarray}
&&F_{x, \alpha \beta}=A_{x,\alpha} + A_{x+\alpha,\beta} - A_{x+\beta,\alpha} - A_{x,\beta},
\\
&&\text{with~~}~ - \pi \le A_{x,\alpha} \le \pi,
\end{eqnarray}
\end{subequations}
which has self-interacting photons.  There is also the non-compact
U(1) QED gauge field theory with the gauge field action
\begin{subequations}
\begin{eqnarray}
&&S=\frac{1}{4g^2}\sum_{x,\alpha \beta} F_{x,\alpha \beta}^2,
\\
&&\text{with}~~~ -\infty  \le A_{x,\alpha} \le  +\infty  ,
\end{eqnarray}
\end{subequations}
which has non-interacting photons.  Even though the compact and the
non-compact theories have the same continuum limit, they have
different confinement properties.  A pair of opposite charges in
non-compact QED interaction are not confined whereas a pair of
opposite charges in compact QED interaction are confined under
appropriate conditions \cite{Pol77,Pol87,Dre79}.
  
As pointed out by Yang \cite{Yan70}, the quantization and the
commensurate property of the electric charges of the interacting
particles imply the compact property of the underlying QED gauge
group.  Because quarks are confined and quark electric charges are
quantized and commensurate, it is reasonable to propose that quarks
and antiquarks interact in the compact QED U(1) interaction.
Accordingly, we can apply the results of Polyakov \cite{Pol77,Pol87},
Drell $et~al.$ \cite{Dre79}, and Schwinger \cite{Sch62,Sch63} to
simplify the dynamics of quarks and the QED gauge field from
(3+1)D$_{\{x^1,x^2,x^3,x^0\}}$ to (1+1)D$_{\{x^3,x^0\}}$, as is
carried out in \cite{Kos21}.  Specifically, Polyakov
\cite{Pol77,Pol87} showed that a pair of opposite electric charges
interacting in compact QED interaction in the transverse
(2+1)D$_{\{x^1,x^2,x^0\}}$ space-time are confined, and that the
transverse confinement persists for all non-vanishing coupling
constants, no matter how weak.  Such a result was subsequently
confirmed by Drell and collaborators \cite{Dre79}.  As explained in
\cite{Dre79}, the transverse confinement of the opposite charges in
compact QED gauge field in (2+1)D${}_{\{x^1,x^2,x^0\}}$ arises from
the periodicity of the gauge field as a function of the spatial
angular variable, leading to transverse gauge photons that are
self-interacting through a periodic and bounded potential in the
neighborhood of the electric charge.  These transverse gauge photons
interact among themselves, they do not radiate away, and they join the
two opposite charges by a confining force \cite{Dre79}.  Upon applying
Polyakov's result to infer transverse confinement of quarks
interacting in compact QED in the transverse space of $\{x^1,x^2\}$,
the dynamics in the (3+1)D$_{\{x^1,x^2,x^3,x^0\}}$ space-time can be
approximated as the dynamics in an idealized\break
(1+1)D$_{\{x^3,x^0\}}$ space-time, with the information in the
transverse degrees of freedom on the $\{x^1,x^2\}$ transverse plane
stored as input parameter properties in the idealized
(1+1)D$_{\{x^3,x^0\}}$ space-time.  It is necessary to examine in this
idealized (1+1)D$_{\{x^3,x^0\}}$ space-time arena whether quarks and
gauge fields are longitudinally confined.  If we next consider further
that the light quarks can be approximated to be massless, then
according to Schwinger's exact solution for massless fermions in such
a (1+1)D$_{\{x^3,x^0\}}$ QED \cite{Sch62,Sch63}, a light quark and its
antiparticle interacting in the QED interaction will be longitudinally
confined just as well and will form a stable QED quark-antiquark
system \cite{Kos21}.
  
From such considerations, as reviewed in Appendix A and B and in
\cite{Won10,Won20,Kos21}, we follow Schwinger and we find
theoretically that there is a stable QED quark current
$j_\qedd^\mu(x)=j_0^\mu(x)$ which obeys the Klein-Gordon equation with
a QED meson mass, $m_{{\qed}\,{ \rm meson}}$, in addition to a stable
QCD quark current $ j_\qcdd^\mu(x)$=$j_1^\mu(x)$ along a unit vector
in the $t^{1,2,...,8}$ generator space which obeys the Klein-Gordon
equation with a different QCD meson mass, $m_{\qcd\,{\rm meson}}$
\cite{Won10,Won20,Won22a,Won22b,Kos21}.  In 1+1 dimensional
space-time, stable collective dynamics of the QED and QCD currents and
their associated gauge fields can be independently excited in their
respective color subgroup spaces.  The corresponding confined QED and
QCD mesons are string-like excitations occurring at different energies
in such 1+1 dimensional space-time.

It is interesting to point out that in previous lattice gauge
calculations, Wilson, t'Hooft, Polyakov, Kogut, Susskind, Mandelstam,
Banks, Jaffe, Peskin,Drell, Guth, Kondo and many others
\cite{Wil74,Pol87,tho74,Kog75,Man75,Ban77,Gli77,Pes78,Dre79,Gut80,Kon98}
showed that fermions and antifermions in the compact QED U(1) gauge
interaction in 3+1 dimensions have a confining phase for strong
coupling and a non-confining phase for weak coupling.  Recent lattice
gauge calculations using tensor networks with dynamical fermions for
compact QED in 3+1 dimensions confirm such a result \cite{Mag20}.  As
quarks are fermions, we expect therefore that under appropriate
conditions, quarks interacting in QED in 3+1 dimensions can be
confined.  The observation and the proposed interpretation of the
anomalous particles as confined composite $q\bar q $ states provide a
special impetus to examine in future lattice gauge calculations
whether quarks in color SU(3) interacting in compact QED U(1) gauge
fields are confined or not, given the quark QED coupling constant and
light quark masses as they are and assuming the QCD gauge fields to be
non-participating spectators.  In this regard, efficient methods of
lattice gauge calculations with dynamical quarks using the tensor
network \cite{Mag20}, dual presentation \cite{Ben20}, magnetic-field
digitization \cite{Bau21}, regulating magnetic fluctuations
\cite{Kap20}, or other efficient methods will be of great interest.
Clearly, whatever the theoretical predictions there may be, the
confining property of quarks interacting in QED interactions in 3+1
dimensions must be tested by experiments.  The present investigation
on QED mesons and the stable quark systems in QED interactions serves
to facilitate the examination of such an important question.

If quarks and antiquarks can interact with the QED interaction alone
under appropriate conditions, can there be other similarly favorable
neutral quark systems stabilized by the QED interaction between the
constituents in the color-singlet subgroup, with the color-octet QCD
gauge interaction as a spectator field?  Of particular interest is the
QED neutron with the $d$, $u$, and $d$ quarks.  The three quarks form
a color product group of\break ${\bb 3}$ $\otimes$ $ {\bb 3} $
$\otimes$ $ {\bb 3}$ = ${\bb 1} \oplus {\bb 8} \oplus {\bb 8} \oplus
{\bb {10}}$, which contains a color-singlet subgroup $\bb 1$ where the
color-singlet current and the color-singlet QED gauge field reside.
In the color-singlet system of the three quarks at energies below the
QCD nucleon mass, the three quarks can interact with the QED
interaction alone.  The QED interaction is attractive between quark
electric charges of opposite signs, and is repulsive between quark
electric charges of the same sign.
\begin{figure} [h]
\centering
\hspace*{0.0cm} \includegraphics[scale=0.70]{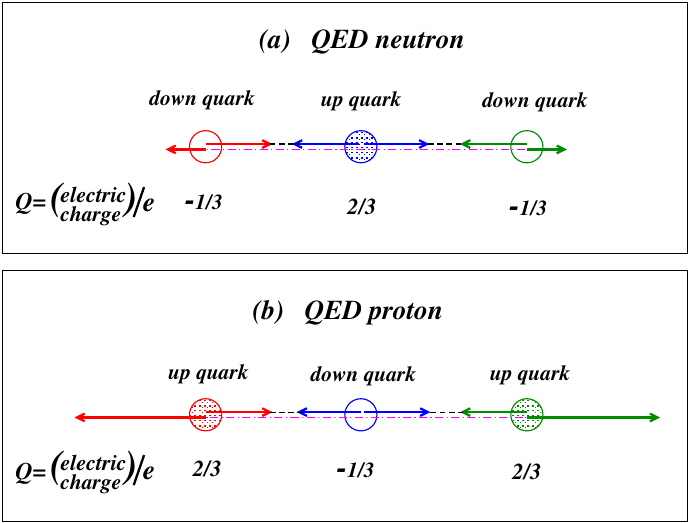}
\caption{The schematic picture of a color-singlet three-quark system
  in 1+1 dimensions in ({\it a}) a QED neutron, and ({\it b}) a QED
  proton.  The lengths of the arrows represent schematically the
  forces acting on each of the quarks arising from the QED
  interactions. The vector sums of the forces acting at each quark
  show that there are attractive forces binding the $d$-$u$-$d$ QED
  neutron and repulsive forces disrupting the $u$-$d$-$u$ QED proton.
}
  \label{eforce}
 \end{figure}
We depict in Fig.\ \ref{eforce} the color-singlet $d$-$u$-$d$ system
with three different colors, and we display the different QED forces
acting on the three quarks, with the magnitude proportional to $|Q_i
Q_j|$ for the force between quark $i$ and quark $j$, an attractive
force for negative $Q_i Q_j$, and a repulsive force for positive $Q_i
Q_j$.  The attractive $d$-$u$ and $u$-$d$ QED interactions between the
two $d$ quarks and the center $u$ quark in Fig.\ \ref{eforce}({\it a})
may overcome the weaker repulsive $d$-$d$ QED interaction between the
two $d$ quarks, to bind the three quarks in the QED neutron.  The
lowest-energy state is likely to reside in 1+1 dimensions because they
are confined, and there is an attraction between the $u$ quark and the
$d$ quarks, but a repulsion between the two $d$ quarks.  However, the
analogous configuration of the QED proton in Fig.\ \ref{eforce}({\it
  b}) with two $u$ quarks and one $d$ quark in the $u$-$d$-$u$
configuration may likely be unstable because of the stronger repulsion
between the two $u$ quarks in comparison with the weaker attractive
interaction between the $d$ and the $u$ quarks.  We would like to show
quantitatively in section 2 that this may indeed be the case.  The
conclusions we will reach concerning the stability of the QED neutron
and proton will also be applicable to the stability of QED
antinucleons.

It is worth pointing out that theoretical and experimental
investigations on the QED neutron are interesting on many accounts.
First is the possibility of its being a new exotic member in the
families of particles of the Standard Model.  Its properties probably
set it apart from other particles because it is a special combination
of the light quark fields and the QED gauge fields that has not yet
been known up to now.  It also calls for a better understanding on the
role of the interplay between different fields in the confinement
process, whether confinement is limited only to constituents in the
presence of the QCD gauge fields, or it suffices to involve also the
light quark fields and the QED gauge fields by themselves without the
QCD interactions.  Clearly, this is a fundamental question that can be
answered only by experiments.  The observation of the anomalous soft
photons, the X17, and the E38 particles indicate that the confined QED
meson states can arise from QED interactions of quarks and antiquarks.
The investigations presented here and elsewhere
\cite{Won10,Won11,Won14,Won20} will provide additional experimental
tests on such a basic property.

There is also the additional interest in the QED neutron, along with
the QED mesons and their corresponding antiparticle counterparts, as
candidate particles for the primordial dark matter because they are
massive, they may be produced during the deconfinement-to-confinement
stage of the quark gluon plasma phase transition, and the quark-gluon
plasma phase may occur in the early history of the Universe or in
high-energy heavy-ion collisions.

To study the stability of the QED neutron and proton with three light
quarks, we need to develop the tools for the relativistic three-body
problem by calibrating the effective two-body QED interaction using
Schwinger's exact QED solution for massless fermions in 1+1 dimensions
in the Appendix.  We generalize the relativistic two-body problem to
the three-body problem and evaluate in section 2 the ground state
energy for the $d$-$u$-$d$ configuration in a variational calculation,
which yields a stable QED neutron.  Section 3 deals with the property
of the lowest state of the QED neutron.  We further examine the
potential energy as a function of the separations between the quarks.
In section 4, we study the stability of the analogous QED $u$-$d$-$u$
proton.  We find that there is no stable energy minimum for the QED
proton.  The total potential between quarks 
interacting in QED alone 
exhibits repulsive
behavior that does not allow the binding of the $u$-$d$-$u$ system.
In section 5, we discuss other favorable quark configurations for
which attractive QED interactions may be present to stabilize the
composite multiquark QED system.  In section 6.  we examine the decay
and the detection of the QED neutron to facilitate its detection.  We
discuss the implication of the instability of the QED proton and the
possibility of the QED neutron to be a dark neutron with a very long
lifetime.  In Section 7, we conclude our discussions.  In Appendix A,
we review Schwinger's exact solution of massless fermions in QED in
1+1 dimensions and see how it may be transcribed into a two-body
problem.  In Appendix B, we study how the color-singlet and the
color-octet excitations arise in the quark-QCD-QED medium by applying
Schwinger's model to massless quarks in QCD and QED in (1+1)D.  In
Appendix C we examine the two-body problem from which the effective
interaction is determined.  Appendix D gives the two-body solution.
Appendix E shows that for the lowest energy two-body bound state a
variational calculation gives the same result as solving the two-body
problem in a wave equation.  Appendix F examines the decays and the
detection of QED mesons.

\section{Stability of the QED  neutron with three quarks  interacting
  in QED interactions}

To study the QED neutron, we construct a composite system of two $d$
quarks and one $u$ quark by selecting quarks of three different colors
to form a color-singlet state and search for its lowest energy bound
state.  In the formulation of Dirac, Todorov, Crater, Van Alstine, and
Sazdjian, and many others
\cite{Dir64,Tod71,Cra83,Cra92,Cra06,Cra07,Cra09,Cra10,Won01a,Saz87,Saz89},
relativistic many-body treatments of bound states have been carried out
in QCD and QED with a high degree of successes.  The basic ingredients
consist of treating particles and antiparticles as independent
positive-energy entities with effective interactions between them.
Each particle obeys a mass-shell constraint on: (i) the momentum, (ii)
the particle mass, and (iii) the effective interactions from the other
particles.  The effective interactions can be obtained by matching
with the perturbative or non-perturbative counterparts of the field
theory or by phenomenological considerations.  In accordance with
Dirac's constraint dynamics \cite{Dir64}, the mass-shell constraints
must however be compatible with each other, resulting in additional
functional requirements or additional terms in the equivalent
Schr\"odinger-type equations whose eigenvalues lead to the eigenstates
and the masses of the composite particle in question.

We consider a three-quark system with a two-body effective interaction
$\Phi _{ij}(x_{ij})$ arising from the particle $j$ at $x_j$ acting on
the particle $i$ at $x_i$, depending on the relative coordinate
$x_{ij}=x_i-x_j$.  The lowest-energy state is likely to reside in 1+1
dimensions because quarks are confined, and there is an attraction
between the $u$ quark and the $d$ quarks but a repulsion between the
two $d$ quarks.  We place the $d$, $u$, and $d$ quarks on the $x$-axis
with coordinate labels $x_1$ and $x_3$ for the two $d$ quarks and
$x_2$ for the $u$ quark.  By allowing all $x_i$ coordinates to assume
both positive and negative values, while fixing the center of mass
position (see Eq.\ (\ref{6p9}) below), we allow all possible
arrangement of the ordering of the positions of the three quarks in
the variations.  We wish to find out quantitatively whether the
attractive QED interactions between the $d$ quarks and the $u$ quark
can overcome the repulsive and weaker QED interaction between the two
$d$ quarks so as to stabilize the QED neutron, as discussed
schematically in Fig.\ \ref{eforce}({\it a}).

For simplicity in our first survey, we neglect particle spins whose
effects on the confining effective interaction are expected to be
small\footnote{Schwinger's exact solution for massless fermions in QED
in 1+1 dimensions indicates that the boson mass of the composite
fermion-antifermion system depends only on the gauge coupling constant
and is independent of the total spin of the fermion-antifermion pair.
Consequently, it can be inferred that spin effects on the effective
confining interaction are small.}.  The spin will lead to a
fine-structure splitting of the QED neutron states which can be
studied when more data become available.  We work in the three-quark
center-of-mass system in which $P=p_1+p_1+p_3$=$(P^0,{\bb P})=(M,0)$,
and the relative coordinate $x_{ij\perp}$=$(x_i-x_j)$
transverse\footnote{ A space-time vector $x_\perp$ is transverse to
another space-time vector $P$ if $x_\perp \cdot P = 0$.}  to $P$
involves only spatial coordinates $x_i$ and $x_j$.  The three particle
momenta in the CM system are
\begin{eqnarray}
p_i &&=(\epsilon_i, q_i), ~~~~ i=1,2,3, \\
\text{where} ~~\epsilon_i&&=\frac{p_i\cdot P}{\sqrt{P^2}},
~~q_i=p_i-\frac{p_i \cdot P } {P^2} P,
\end{eqnarray}
and we consider the particles to be of positive energy only, with
$\epsilon_i>0$.  The rest mass $M$ of the composite particle is
 \begin{eqnarray}
M=P^0=\epsilon_1+\epsilon_2+ \epsilon_3.
\end{eqnarray}
We generalize the two-body equations of (\ref{3p1a}) and (\ref{3p1b})
to the three-body problem by imposing three mass-shell constraints
relating the momenta, the masses, and their interactions in the form
\begin{subequations}
\label{6p3}
\begin{eqnarray}
{\cal H}_1|\Psi\rangle =\biggl \{ p_1^2 - m_1^2 - [ \Phi_{12} (x_{12}) +\Phi_{13} (x_{13}) ]  \biggr \} |\Psi\rangle=0,
\label{6p3a}\nonumber\\
\\
{\cal H}_2|\Psi\rangle=\biggl \{ p_2^2 - m_2^2 - [ \Phi_{21} (x_{21}) +\Phi_{23} (x_{23}) ]  \biggr \} |\Psi\rangle=0,
\label{6p3b}\nonumber\\
\\
{\cal H}_3|\Psi\rangle=\biggl \{    p_3^2 - m_3^2 - [ \Phi_{31} (x_{31}) +\Phi_{32} (x_{32}) ]  \biggr \} |\Psi\rangle=0.
\label{6p3c}\nonumber\\
\end{eqnarray}
\end{subequations}
The compatibility conditions on the above mass-shell constraints lead
to the requirement that $\Phi_{ij}(x_{ij})$=$\Phi_{ji}(x_{ij})$ and
the variable $x_{ij}$ in the effective interaction $\Phi_{ij}(x_{ij})$
be $x_{ij}=x_{ij\perp}'$, which is transverse to
the combined momentum $P_{ij}$=$p_i+p_j$.  This $x_{ij \perp}'$
coordinate should be the relative spatial coordinate in the frame in
which the center-of-mass of the system of constituents $i$ and $j$ is
at rest.  For the three-body problem, the center-of-mass motion of any
two constituents $i$ and $j$ has a velocity
$V_{ij}$=$(q_i+q_j)/(\epsilon_i + \epsilon_j)$ and may not be at rest.
Even though $V_{ij}$ may not be zero, it will be constrained and
limited in a bound state. It is reasonable to neglect such velocities
so that we can approximate the $x_{ij \perp}'$ that is transverse to
momentum $P_{ij}$ to be the relative coordinate $x_{ij \perp}$ that is
transverse to the total center-of-mass momentum $P$ instead.  In such
an approximation, the relative coordinate $x_{ij \perp}'$ in the
effective interaction $\Phi_{ij}(x_{ij\perp}')$ becomes just the
relative coordinate $x_{ij\perp}$ of $i$ and $j$ in the three-quark
center-of-mass system.  Corrections for such an approximation will be
of order $V_{ij}$, which can be taken as velocity-dependent
corrections in future refinements.

In the wave equations (\ref{6p3a})-(\ref{6p3c}), the phenomenological
effective QED interaction extracted from\break Schwinger's exact QED
solution in Eqs.\ (\ref{4p1}) and (\ref{4p9}) in Appendix D is
\begin{eqnarray}
\Phi _{ij}(x_{ij\perp})\!=\!\frac{2\epsilon_i \epsilon_j}{\epsilon_i+\epsilon_j} (- Q_i Q_j )\kappa |x_i-x_j|  ~\text{with}~\kappa\!=\!\frac{e^2}{4\pi},
\label{6p4}
\end{eqnarray}
where $e$=$e_\2d=g^\qedu_\2d$ is the QED coupling constant\footnote{We
adopt here the notations that $e$ is actually $e_\2d$, the QED
coupling constant in 1+1 dimensions, and $e_\4d$ is the QED coupling
constants in 3+1 dimensions, with the fine structure constant defined
by $\alpha$=$\alpha_\4d$=$e_\4d^2/4\pi=1/137$.  As pointed out in
\cite{Won09,Won10,Kos12,Kos21}, $e_\2d$ and $e_\4d$ are related by the
flux tube radius $R_T$ as $e_\2d^2=e_\4d^2/(\pi R_T^2)$, when the
confining flux tube is approximated as an open string without a
structure.  }  in (1+1)D.

Equations (\ref{6p3}) and (\ref{6p4}) constitute the system of
relativistic three-body equations for three quarks interaction in the
effective QED interactions.  We would like to investigate whether
there is a lowest-energy equilibrium state of the QED neutron by using
a variational wave function.  It is convenient to choose a Gaussian
variational wave function of the spatial dimensionless spatial
variables $y_1, y_2, y_3$ with standard deviations $\sigma_1$,
$\sigma_2$, and $\sigma_3$ as variational parameters,
\begin{eqnarray}
\Psi(y_1,y_2,y_3)= N\exp \biggl \{ -\frac{y_1^2}{4\sigma_1^2}
-\frac{y_2^2}{4\sigma_2^2}-\frac{y_3^2}{4\sigma_3^2} \biggr \},   
\label{15}
\end{eqnarray}
where $y_i=\sqrt{\kappa}x_i$.  The charge numbers of the quarks are
$Q_1=Q_3=-1/3$, and $Q_2 =2/3$.  The expectation values of
(\ref{6p3a})-(\ref{6p3c}) using the variational wave function $\Psi$
are
\begin{subequations}
\label{6p6}
\begin{eqnarray}
&&\!\!\!\!\langle \Psi |  \frac{\epsilon_1^2}{\kappa} | \Psi  \rangle \!=\! \langle \Psi |\! \biggl \{  
\biggl[\frac{1}{2\sigma_1^2} -\frac{y_1^2}{4\sigma_i^4}\biggr ]
\!+\!\frac{m_1^2 }{\kappa}
\nonumber\\
&&\!+\! 
\frac{2\epsilon_1 \epsilon_2}{\epsilon_1+\epsilon_2} (\frac{2}{9}) |y_1-y_2|
\!+\!\frac{2\epsilon_1 \epsilon_3}{\epsilon_1+\epsilon_3} 
(\frac{-1}{9}) |y_1-y_3|
 \biggr \}\! |\Psi \rangle,~~~~~~~
\label{6p6a}
\\
&&\!\!\!\!\langle \Psi | \frac{\epsilon_2^2}{\kappa}  | \Psi  \rangle \!=\! \langle \Psi | \biggl  \{  
\biggl[\frac{1}{2\sigma_2^2} -\frac{y_2^2}{4\sigma_2^4}\biggr ]
\!+\!\frac{m_2^2 }{\kappa}
\nonumber\\
&&\!+\! 
\frac{2\epsilon_2 \epsilon_1}{\epsilon_2+\epsilon_1} 
(\frac{2}{9})
|y_2-y_1|
\!+\!\frac{2\epsilon_2 \epsilon_3}{\epsilon_2+\epsilon_3} 
(\frac{2}{9})
 |y_2-y_3|
 \biggr \}\! |\Psi \rangle,~~~~~~~~
\\
&&\!\!\!\!\langle \Psi |  \frac{\epsilon_3^2}{\kappa}  | \Psi  \rangle\!=\! \langle \Psi | \!\biggl \{  \!
\biggl[\frac{1}{2\sigma_3^2} -\frac{y_3^2}{4\sigma_3^4}\biggr ]
\!+\!\frac{m_3^2 }{\kappa}
\nonumber\\
&&\!+\!\frac{2\epsilon_3 \epsilon_1}{\epsilon_3+\epsilon_1} 
(\frac{-1}{9}) |y_3-y_1| 
\!+\!\frac{2\epsilon_3 \epsilon_2}{\epsilon_3+\epsilon_2} 
(\frac{2}{9}) |y_3-y_2|
 \biggr \} \!|\Psi \rangle.~~~~~~~
\label{6p6c}
\end{eqnarray}
\end{subequations}
Because of the symmetry of the two $d$ quarks we can assume for the
lowest-energy state
\begin{eqnarray}
\sigma_1=\sigma_3,
\end{eqnarray}
so that the variational parameters consist only of $\sigma_1$ and
$\sigma_2$.  We look for the state with the lowest composite mass $M$
in the variations of $\sigma_1$ and $\sigma_2$,
\begin{eqnarray}
\frac{\delta^2 M(\sigma_1,\sigma_2)}{\delta \sigma_1 \delta \sigma_2}
= 0.
\label{5p7}
\end{eqnarray}
The motion of the three quarks should maintain a fixed center of mass
for the composite system.  It is necessary for the coordinates of the
three quarks to satisfy the center-of-mass condition on the spatial
coordinates,
\begin{eqnarray}
\sum_{i=1}^3  \epsilon_i  y_i=0.
\label{6p9}
\end{eqnarray}
The variational wave function $\Psi$ of Eq.\ (\ref{15}) is normalized according to
\begin{eqnarray}
\!\!\int \!\! dy_1 dy_2 dy_3 |\Psi(y_1,y_2,y_3)|^2 \delta (\epsilon_1 y_1+\epsilon_2 y_2+ \epsilon_3 y_3) \!=\! 1.
\end{eqnarray}
Because of the CM condition, there are actually only two independent
spatial variables which can be chosen to be $y_1$ and $y_3$.  However,
we need to treat all three spatial variables as independent in the
beginning, and impose the CM constraint (\ref{6p9}) only when we
evaluate the expectation values in (\ref{6p6}) to calculate
$\epsilon_i$ and $M$ at the end.

\begin{figure} [h]
\centering
\includegraphics[scale=0.40]{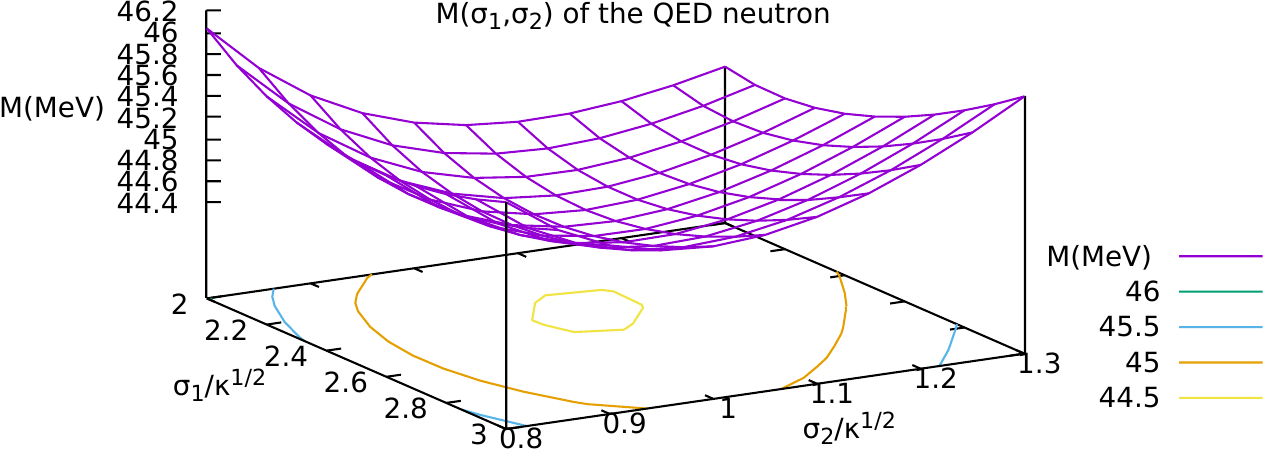}
\vspace*{0.3cm}
\caption{The mass $M$ of the QED neutron as a function of the
  variation parameters $\sigma_1, \sigma_2$ in units of
  $\hbar/\sqrt{\kappa}$=8.29 fm.  The QED neutron has an energy
  minimum at $M=44.5$ MeV at $\sigma_1/\sqrt{\kappa}=2.40$ and
  $\sigma_2/\sqrt{\kappa}=1.09$.}
\label{figmin}
\vspace*{0.2cm}
\end{figure}
 
In the evaluation of the QED neutron mass $M$, the unknown quantities
$\epsilon_i$ are needed to defined the effective interactions.  They
can be obtained self-consistently and iteratively with initial
guesses.  Knowing the effective interactions and the given variational
parameters $\sigma_1$ and $\sigma_2$, we evaluate the expectation
values on the right hand sides of (\ref{6p6a})-(\ref{6p6c})
numerically.  The calculated values of $\epsilon_i$ on the left hand
sides of (\ref{6p6a})-(\ref{6p6c}) can form the basis of the next
iteration until convergence is achieved.  In the numerical
calculations, we have used quark masses $m_u=2.16$ MeV and $m_d=4.67$
MeV \cite{PDG19}.

By such variational calculations, we find that the mass $M$ as a
function of $\sigma_1$ and $\sigma_2$ has an energy minimum, $M=44.5$
MeV, at $\sigma_1=2.40\hbar/ \sqrt{\kappa}$=19.9 fm and
$\sigma_2=1.05\hbar /\sqrt{\kappa}$=8.71 fm as shown in
Fig.\ \ref{figmin}.
\vspace*{0.6cm}

\section{Properties of the lowest-energy QED neutron}

Table 1 lists the physical properties of the QED neutron at its energy
minimum.  The two $d$ quark energies $\epsilon_1$ and $\epsilon_3$ are
smaller than the $u$ quark energy $\epsilon_2$ because the two
effective interactions in Eq.\ (\ref{6p6}) for each of the $d$ quarks
have opposite signs while those for the $u$ quark have the same sign.
The root-men-squared separation between the two $d$ quarks is 28.2 fm
and thus the QED neutron spans a length of order many tens of fermi.
The wave function of the $d$ quarks have a larger value of the
standard deviation $\sigma_1$ as compared to the standard deviation
$\sigma_2$ of the $u$ quark.  In the classical description of an open
string \cite{Art74,And83}, the $d$ quarks shuttle about the central
$u$ quark in a yo-yo motion back and forth from the left to the right
of the $u$ quark and back.  When a $d$ quark comes to one side of the
$u$ quark, the other $d$ quark goes to the other side to balance the
center-of-mass motion.  The $u$ quark itself also makes excursions
about the geometrical center, as indicated by a smaller value of the
standard deviation $\sigma_2$.
\begin{table}[h]
\centering
\caption{Properties of the lowest-energy QED neutron}
\begin{tabular}{|l | c |}
\hline
~~~~Quantity   & Value  \\
\hline
 ~~~~$M$ (mass of the QED neutron) & 44.5 MeV    \\ 
  $\sqrt{\langle \epsilon_1^2 \rangle }$=$\sqrt{\langle \epsilon_3^2 \rangle }$~~(the $d$ quark)  & 11.3 MeV \\
  $\sqrt{\langle \epsilon_2^2 \rangle }$ ~(the $u$ quark)~~  & 21.8 MeV \\
$\sqrt{\langle (x_1\!-\!x_2)^2 \rangle}$(between $d$ quark and $u$ quark) & 20.4 fm  \\
 $\sqrt{\langle (x_3-x_1)^2 \rangle }$ (between two $d$ quarks) & 28.2 fm  \\
 $\sigma_1$ (of wave function for the $d$ quarks)   & 19.9 fm   \\ 
 $\sigma_2$ (of wave function for the $u$ quark)   & ~8.71 fm   \\ 
\hline
\end{tabular}
\end{table}

The force vectors in Fig.\ \ref{eforce}({\it a}) give a qualitative
description of the various forces leading to the binding of the three
quarks in a QED neutron.  The variational calculations demonstrate the
stability of the QED neutron in a quantitative analysis.  It is
illuminating to see how the effective interactions between the three
quarks can bind them together into a QED neutron from a more
quantitative viewpoint as an illustration.  For such a purpose, we add
the wave equations in (\ref{6p3}) and we get the total mass-shell
condition
\begin{eqnarray}
\biggl \{ && \!\!\sum_{i=1}^3(\epsilon_i^2 -m_i^2)    -\sum_{i=1}^3 \bb q_i ^2 
\\
&&\hspace*{-0.3cm}-2[ \Phi_{12} (x_{12}) +\Phi_{13} (x_{13}) +\Phi_{23} (x_{23})]\biggr \} \Psi(x_1,x_2,x_3)\!=\!0.
\nonumber
\end{eqnarray}
This is just  a three body system with a total effective interaction 
\begin{eqnarray}
\Phi_{\rm tot}(x_1,x_2,x_3)\!=\!2[ \Phi_{12} (x_{12}) \!+\!\Phi_{13} (x_{13}) \!+\!\Phi_{23} (x_{23})] .
\end{eqnarray}
\begin{figure} [h]
\centering \includegraphics[scale=0.45]{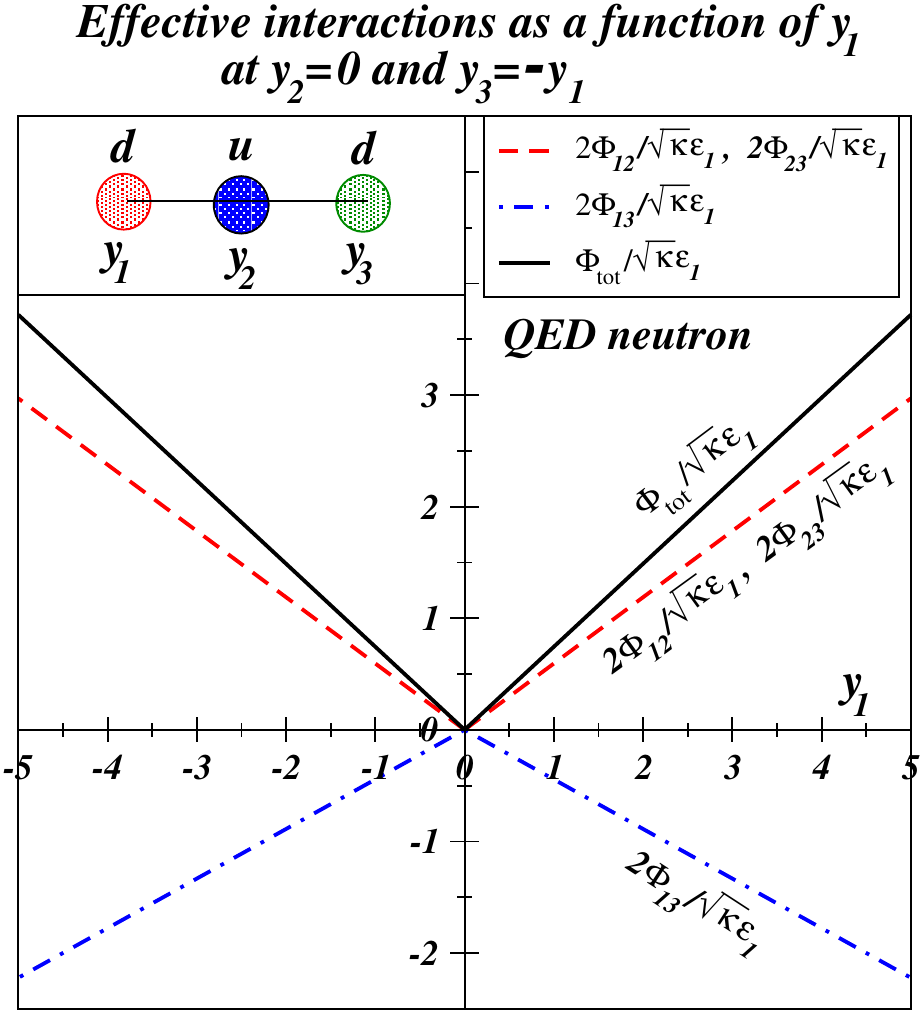}
\caption{ The effective interaction $2\Phi_{12}$, $2\Phi_{23}$,
  $2\Phi_{31}$, and $\Phi_{\rm tot}=2\Phi_{12}+2\Phi_{23}+2\Phi_{31}$
  in units of $\sqrt{\kappa }\epsilon_1$ for a QED neutron, where
  $\Phi_{ij}$ is the effective interaction between quarks at $y_i$ and
  $y_j$ and $\Phi_{\rm tot}$ is the total effective interaction, for a
  selected sample $d$-$u$-$d$ arrangement of the three quarks shown at
  the upper left corner.  The potentials are obtained as a function of
  $y_1$, at $y_2=0$, $y_3=-y_1$, where $y_1$, $y_2$ and $y_3$ are the
  positions of the $d$, $u$, and $d$ quarks respectively, }
\label{fign}
\end{figure}
We can acquire a better understanding how the three quarks can bind
together in the QED neutron when we examine various components of the
effective interactions between different pairs of quarks as a function
of a set of representative spatial coordinates.  We can choose a
sample set of representative coordinates such that the first $d$ quark
coordinate is $y_1$, the $u$ quark coordinate is at $y_2$=0, and the
second $d$ quark coordinate is at $y_3$=$-y_1$ because of the CM
constraint.  For such a sample set of representative coordinates, we
can study the behavior of various effective interactions which can be
expressed as functions of a single variable $y_1$.  The various
effective interactions $2\Phi_{ij}/\sqrt{\kappa}\epsilon_1$ between
quark $i$ at $y_i$ and quark $j$ at $y_j$ and the total $\Phi_{\rm
  tot}$ as a function of $y_1$ are
\begin{eqnarray}
&&\!\!\!\!\!\frac{2\Phi_{12} (y_1,y_2)}{\epsilon_1 \sqrt{\kappa}}\!\biggr |_{y_2=0,y_3=-y_1} \hspace*{-0.8cm}
=\!\frac{2\Phi_{23} (x_{23})}{\epsilon_1 \sqrt{\kappa}} \!\biggr |_{y_2=0,y_3=-y_1}
\hspace*{-0.6cm}=\!\frac{4\epsilon_2/\epsilon_1}{1\!+\!\epsilon_2/\epsilon_1} \!\frac{2}{9}\! |y_1|,
\nonumber\\
&&\!\!\!\frac{2\Phi_{13} (y_1,y_3)}{\epsilon_1 \sqrt{\kappa}} \biggr |_{y_2=0,y_3=-y_1}\hspace*{-0.4cm}
=\frac{8\epsilon_3/\epsilon_1}{1+\epsilon_3/\epsilon_1} (\frac{-1}{9}) |y_1|\!=\!(\frac{-4}{9}) |y_1|,
\nonumber\\
&&\!\!\frac{\Phi_{\rm tot}(y_1,y_2,y_3)}{\epsilon_1 \sqrt{\kappa}}\biggr |_{y_2=0,y_3=-y_1}\hspace*{-0.6cm}
=\left \{\! \frac{8\epsilon_2/\epsilon_1}{1+\epsilon_2/\epsilon_1} (\frac{2}{9})  + (\frac{-4}{9}) \!\right \}\! |y_1|.
\nonumber
\end{eqnarray}
At the minimum energy point, the values of $\epsilon_1$ and
$\epsilon_2$ are $\epsilon_1\!=\!0.4767\sqrt{\kappa}$,
$\epsilon_2\!=\!0.9163\sqrt{\kappa}$, and so
$\epsilon_2/\epsilon_1\!=\!1.922$ and the above dependencies can be
evaluated.  We show the effective interactions $\Phi_{\rm tot}$ and
$\Phi_{ij}$ between different quarks, as a function of $y_1$ for the
QED neutron at $y_2=0$, $y_3=-y_1$ in Fig.\ \ref{fign}.  The
attractive $u$-$d$ interactions $2\Phi_{12}/\sqrt{\kappa}\epsilon_1$
and $2\Phi_{23}/\sqrt{\kappa}\epsilon_1$ are shown as the dashed
curve.  The repulsive $d$-$d$ interaction
$2\Phi_{13}/\sqrt{\kappa}\epsilon_1$ is shown as the dashed-dot curve
in Fig.\ \ref{fign}.  The total effective interaction $\Phi_{\rm tot}$
is displayed as the solid curve which is a confining interaction that
binds the three quarks together.  Hence there is a stable QED neutron
arising from the balances of the mutual electrostatic forces between
the quarks.

With the above confining potential in Fig.\ \ref{fign} as an
illustrative example, we can understand the energy minimum of the QED
neutron in two intuitive ways.  In the description of the classical
string \cite{Art74,And83}, the two $d$ quarks execute yo-yo motion
shuttling about the $u$ quark back and forth after reaching the
longitudinal turning points in the confining potential of
Fig.\ \ref{fign}.  In the quantum mechanical description, the
attractive net confining QED interaction as shown in Fig.\ \ref{fign}
between the quarks is counterbalanced by the quantum stress pressure
\cite{Won76} that arises from the derivatives of the single-particle
wave function, reaching the lowest-energy equilibrium between the
attractive QED interaction and the quantum stress pressure.

\section{The stability of the  QED proton and the  QED neutron weak decay} 

We would like to study next whether QED color-singlet proton with two
$u$ quarks and a $d$ quark can be stable.  For such a calculation, we
carry out the variational calculations as in the above QED neutron
case, with the $u$ and $d$ quarks in the QED neutron replaced by $d$
and $u$ quarks respectively.  That is, we consider the $u$, $d$, and
$u$ quarks to be placed on the $x$-axis with coordinate labels $x_1$
and $x_3$ for the two $u$ quarks and $x_2$ for the $d$ quark.  By
allowing all $x_i$ coordinates to assume both positive and negative
values, while fixing the center of mass position (Eq.\ (\ref{6p9})),
we allow all possible arrangement of the ordering of the positions of
the three quarks in the variations, including both the linear
$u$-$d$-$u$ configuration as in Fig.\ \ref{eforce}({\it b}) and the
$u$-$u$-$d$.  In this case of QED proton, we have $Q_1=2/3$,
$Q_2=-1/3$, and $Q_3=2/3$.

\begin{figure} [h]
\centering \includegraphics[scale=0.45]{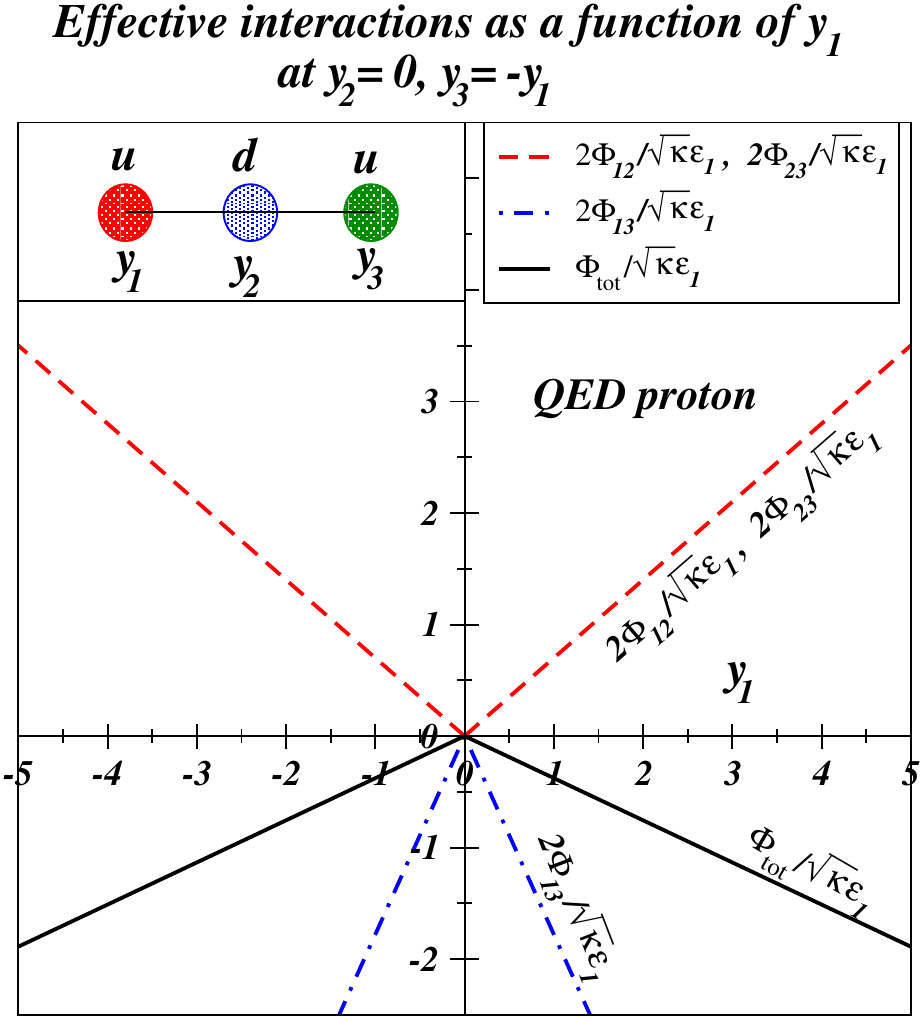}
\caption{ The effective interaction $2\Phi_{12}$, $2\Phi_{23}$,
  $2\Phi_{31}$, and $\Phi_{\rm tot}=2\Phi_{12}+2\Phi_{23}+2\Phi_{31}$
  in units of $\sqrt{\kappa }\epsilon_1$ for a QED proton, where
  $\Phi_{ij}$ is   the effective interaction 
  between quarks at $y_i$ and $y_j$ and $\Phi_{\rm
    tot}$ is the total effective interaction, for a selected sample
  $u$-$d$-$u$ arrangement of the three quarks shown at the left upper
  corner.  The potentials are obtained as a function of $y_1$, at
  $y_2=0$, $y_3=-y_1$, where $y_1$, $y_2$ and $y_3$ are the positions
  of the $u$, $d$, and $u$ quarks respectively, The total effective
  interaction $\Phi_{\rm tot}$ shown as the solid curve is a linear
  repulsive interaction, indicating that the QED proton is not stable.
}
\label{figp}
\end{figure}

Our variations over a very large range of $\sigma_1$ and $\sigma_2$
values fail to find an energy minimum.  Extending the range of
$\sigma$ will only drive the total energy of the system lower with the
$u$ quarks farther and farther apart without the energy turning to a
minimum.  The condition of (\ref{5p7}) cannot be satisfied for this
case.  We can understand the failure by looking at the total potential
2$\Phi_{\rm tot}$ at a sample arrangement shown in Fig.\ \ref{figp}.
The effective potentials at $y_2=0$, $y_3=-y_1$ for the sample case
with $\sigma_1=\sigma_3=2.09\sqrt{\kappa}$ and
$\sigma_2=2.80\sqrt{\kappa}$, which give $\epsilon_2/\epsilon_1=3.71$
are shown in Fig.\ \ref{figp}.  The magnitude of the sum of the
attractive effective interactions 2$\Phi_{12}$=2$\Phi_{23}$ between
the $d$ quark and the two $u$ quarks is smaller than the magnitude of
the repulsive interaction 2$\Phi_{13}$ between the two $u$ quarks.
The total effective interaction $\Phi_{\rm tot}$ is repulsive; it
decreases as $|y_1|$ increases.  Hence, a QED proton does not possess
a stable bound state.  The QED proton also does not possess a
continuum state with isolated quarks because the isolation of quarks
as color-triplet quarks is forbidden.  Therefore, the QED proton does
not exist either as a stable bound state nor a continuum state with
isolated quarks.  There is no QED proton state.
 
The absence of a QED proton state has an important consequence on the
weak decay of the QED neutron.  The weak decay of the QED neutron
occurs when a $d$ quark in the three-quark system decays into an $u$
quark.  Such a QED neutron weak decay would result in a possible QED
proton final state, if a QED proton state could exist.  Because there
is no final bound or continuum QED proton state for the QED neutron to
decay onto, the density of final states for the weak decay of a QED
neutron onto a QED proton is zero.  Consequently the rate of the QED
weak decay into a QED proton is zero.  The QED neutron can only decay
by a baryon-number non-conserving transition which presumably has a
very long life time.  Therefore, the lowest energy QED neutron is a
stable particle with a very long lifetime and is in fact a dark
neutron.

\section{Other favorable QED quark configurations and QED quark matter}

\begin{figure} [h]
\centering
\includegraphics[scale=0.80]{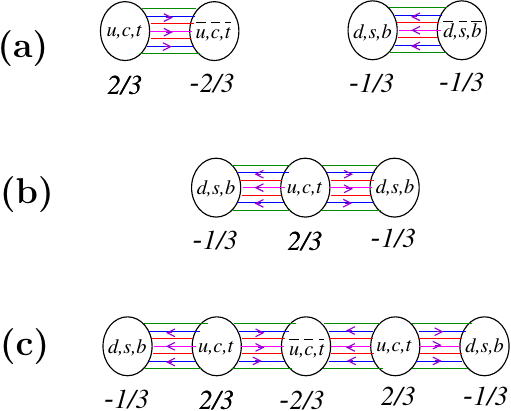} 
\vspace*{0.30cm}
\caption{ Some favorable neutral open-string configurations for the
  lowest energy states involving (a) a quark and an antiquark, (b)
  three quarks, and (c) three quarks plus a quark-antiquark pair, for
  quarks and antiquarks interacting in QED forces in the color-singlet
  $\bb 1$ subgroup of the product group.  Their electric field lines
  of force are indicated by arrows and the quark electric charge
  numbers are listed.  These open-string configurations may be
  stabilized by the linear QED forces which are attractive between
  charges of opposite signs and repulsive between charges of the same
  sign.  }
\label{others}
\end{figure}

Although the QED proton with two up quarks and a down quark cannot
have a stable bound or continuum state, other neutral QED quark and
antiquark systems and their corresponding charge-conjugate
counterparts may have stable configurations arising from quarks and
antiquarks interacting in QED interactions in the color-singlet $\bb
1$ subgroup of the direct product color group, with the QCD gauge
interactions as spectator fields.  Whether heavy quarks can also
interact in QED interactions in this sector cannot be excluded, and it
is of interest to examine such a possibility theoretically so as to
facilitate future experimental assessments on the occurrence
probabilities.  By including heavy quarks also among the quarks
interacting in QED interactions with the QCD gauge fields as
non-participating spectators in Fig.\ \ref{others}, there can be a
large number of favorable configurations of neutral open-string
systems in which the forces acting on the quarks and antiquarks are
subject to a linear QED force that is attractive between two charges
of opposite signs and repulsive for two charges of the same sign.  The
QED attractive forces in these configurations may overwhelm the
repulsive forces to lead to stable color-singlet QED composite
particles with different flavor contents and quark numbers.

In Fig.\ \ref{others}, we list different choices of flavors for each
of the quarks.  The quark and the antiquark in Fig.\ \ref{others}(a)
have electric charges of opposite signs, and the attractive QED forces
between them can stabilize the system.  There can be additional flavor
mixing considerations if one further assumes flavor SU(2) or SU(3)
symmetries as discussed in \cite{Won20}.  Fig.\ \ref{others}(b) shows
three quarks, with two quarks of electric charges of $(-1/3)$ and a
quark of charge (2/3).  They are the analogue of the QED neutron and
are likely to be stable.  Fig.\ \ref{others}(c) shows a linear chain
of five quarks.  An example of such a configuration is $d$-$u$-$\bar
u$-$u$-$d$ which can be built even longer as $d$-$u$-($u$-${\bar
  u})^n$-$d$, with $n=0,1,2..$.  It have electric charges with
alternating signs such that the QED forces between them may be
attractive and balanced to stabilize the system.

In the configurations in Fig.\ \ref{others}, one can construct
color-singlet states by choosing color-anticolor combinations for the
quark-antiquark combinations in Fig.\ \ref{others}(a), three different
colors for the three-quark combination in Fig.\ \ref{others}(b), and
three different colors for the three-quark combination, and a color
and an anticolor for the additional quark-antiquark pair in
Fig.\ \ref{others}(c). It will be of interest to study theoretically
and experimentally whether these neutral color-singlet quark systems
are stable.

Stable QED mesons and QED neutrons, if found to occur, can lead to the
QED composite matter, a new type of matter that is a collection of QED
mesons and QED neutrons interacting with each other by weak Van der
Waal-type interactions between electric multipoles.  The QED composite
matter can make a transition to become a deconfined quark-QED plasma
at the phase transition temperatures $T_{\rm trans}^\qedu$.  By
dimensional analysis following Eq.\ (\ref{eq1}), the ratio of the
phase-transition temperatures, $T_{\rm trans}^{\qedu}$/ $T_{\rm
  trans}^{\qcdu}$, at which QED and QCD mesons become deconfined,
would be approximately proportional to their gauge field coupling
constants.  This places $T_{\rm trans}^{\qedu}$$\sim$ $T_{\rm
  trans}^{\qcdu}$/10 $\sim$ 20 MeV.  In the finite baryon density
regime, the deconfinement phase transition occurs when the degenerate
fermion pressure of the dense QED neutrons overwhelming the QED
confinement interaction \cite{Won94}.  The quark-QED plasma with
varying net baryon densities will have its equilibrium density and its
own equation of state.  They may be produced in high-energy heavy ion
collisions, in the core of neutron stars, or in some stages of neutron
star mergers.  It has been suggested that the deconfined phase of
quark matter may be present in the core of massive neutron stars
\cite{Ann20}.  In this case, QED quark photon plasma and QED neutrons
may also be present in the hadron-quark matter transition region close
to the deconfined quark matter core.  Future studies on the quark-QED
plasma properties and its possible phase transition to QED neutrons
will be of great interest.

\section{The detection of a QED neutron}

A QED neutron is a composite object containing two $d$ quarks and one
$u$ quark interacting in QED interactions.  The relativistic
three-body equations of (\ref{6p3a}), (\ref{6p3b}), (\ref{6p3c}) with
a confining potential such as shown in Fig.\ \ref{fign} are expected
to have many eigenstates and eigenvalues.  The inclusion of spin-spin,
spin-orbit, and other interactions between the quarks in a future
fully three-dimensional calculation will add a greater degree of
complexity to the spectrum of the QED neutron.

As the detection of a QED neutron will depend on its decay products,
we would like to examine how a QED neutron decays from an excited
state $n_{\rm QED}^*$(initial) to a final state $n_{\rm QED}'$(final).
We note first of all what it will not do.  It will not dissociate
itself into isolated quark constituents because of the non-isolation
nature of the quarks.  It will not decay by weak interactions into a
positively charged entity because there is no bound or continuum final
QED proton state.

\begin{figure} [h]
\centering \includegraphics[scale=0.95]{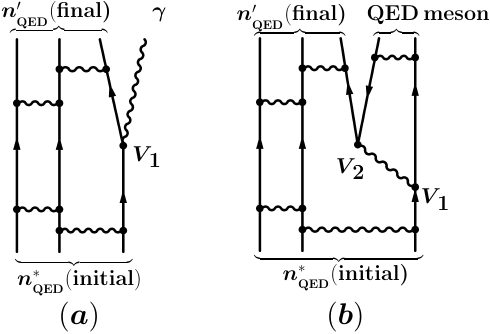} 
\caption{ The decay modes of a QED neutron from an excited eigenstate
  $n_{\rm QED}^*$(initial) to the final eigenstate $n_{\rm
    QED}'$(final): ({\it a}) by the emission of a photon, and ({\it
    b}) by the emission of a QED meson which subsequently can decay
  into real or virtual photons, or a dilepton pair as described in
  Fig.\ \ref{decay}(a), \ref{decay}(b), and \ref{decay}(c) in Appendix
  F.  }
\label{ndecay}
\end{figure}

In 1+1 dimensions, an excited state of QED neutron cannot decay as the
photon is represented in effect by an effective potential $\Phi$ as
discussed in Appendix C, and the quarks do not radiate photon. In the
physical 3+1 dimensions, the transverse structure of the flux tube
must be taken into account and the photon emission channel from the
quark opens up.  A quark can make a sharp change of its trajectory
turning to the transverse direction with the emission of a photon from
the excited $n_{\rm QED}^*$.  By such an emission process at the
vertex $V_1$ as depicted in diagrams in Fig.\ \ref{ndecay}(a), a quark
in an excited QED neutron at the initial excited eigenstate
$n_\qedd^*$(initial) can de-excite to reach the final eigenstate
$n_\qed'$(final).  The multipolarity of the photon transition will
depend on the spins and the parities of the initial and final states
in question.  Alternatively, a valence quark in an initial excited
eigenstate $n_\qedd^*$(initial) can de-excite by the emission of a
photon at the vertex $V_1$ in Fig.\ \ref{ndecay}(b) leading to the
production of a quark-antiquark pair at the vertex $V_2$.  The
produced quark can join up with the remaining two quarks of the
initial QED neutron $n_\qed^*$(initial) to become the final QED
neutron eigenstate $n_\qed'$(final), while the produced antiquark can
combine with the valence quark to form a QED meson.  In such a decay
with the emission of a QED meson as depicted in
Fig.\ \ref{ndecay}({\it b}), the flavor of the produced pair at $V_2$
must agree with the flavor of the valence quark emitting the virtual
photon at $V_1$ so that the flavor and the charge of the final QED
neutron remains unchanged.

It is easy to envisage that successive emissions of photons and QED
mesons will allow an excited QED neutron $n_{\rm QED}^*$ to de-excite,
and eventually to reach the lowest energy QED neutron state.  Through
out the de-excitation process, the three quarks constituents remain
bound to each other as an entity, in the conservation of the QED
neutron number, while the emitted QED mesons will decay into real
photons, virtual photons in the form of dileptons, or a dilepton pair
as described by diagrams Figs.\ \ref{decay}(a), \ref{decay}(b), and
\ref{decay}(c) in Appendix F.  At the end of the de-excitation, only
the ground state QED neutron remains and it does not radiate because
it is the lowest energy QED neutron state.  Because the lowest-energy
QED neutron ground state does not decay or radiate, it is therefore a
QED dark neutron.

With the exception of the QED dark neutron which has no decay
products, the detection of a QED neutron can be carried out by
searching for their decay products of photons and QED mesons arising
from the de-excitation of the excited QED neutron states with the
emitted QED mesons detected as diphoton resonances.  We envisage that
by the coalescence of the quarks of different colors, QED neutrons at
the lowest-energy state as well as the excited states may be produced
during the deconfinement-to-confinement phase transition of the quark
gluon plasma. The de-excitation of the excited QED neutron states will
yield photons and QED mesons of various energies exhibiting the
spectrum of the QED neutron system.  The de-excitation energies
provide information on the QED neutron structure.  We may rely on the
presence of these emitted photons and QED mesons to reconstruct the
complete spectrum of the QED neutron.  The de-excitation may also go
through many steps with sequential emissions of QED mesons and/or
photons.  Accordingly, we can look for unknown photons and QED mesons
that accompany the production of other photons and QED mesons.

We note with keen interest that QED neutrons may be produced by the
coalescence of deconfined quarks in the core of a dense neutron star
or in high-energy heavy-ion collisions during the phase transition of
the quark gluon plasma.  The production of the QED neutron or its
excited states can be used as a signature for the quark-gluon plasma
production, if the QED neutron can be so identified.  It is necessary
to identify the QED neutron if it is produced.  A QED neutron can be
identified by its QED meson emission spectral lines which exhibit its
own characteristic structure.  With the mass of the lowest energy QED
neutron state predicted to be 44.5 MeV, and if a harmonic oscillator
spectrum for the QED neutron can be an order-of magnitude guide, we
expect the masses of the emitted QED mesons in the decay of an excited
QED neutron state to be of order 50 MeV.  We can therefore estimate
the production of diphoton resonances with an invariant mass of order
50 MeV to accompany the production of the QED neutron.  A way to
distinguish those QED mesons as arising from the QED neutron decay or
from a $q\bar q$ pair production is to study the differences of the
QED meson emission spectral lines as a function of the probability for
the occurrence of deconfinement, which is correlated with the mass,
energy, and centrality of the high-energy collision process.  For
those collisions with a low probability leading to deconfinement, the
detected QED mesons emission spectral lines will not contain those
lines emitted from the excited QED neutrons.  On the other hand, for
collisions with a high likelihood of attaining deconfinement, there
will be excited QED neutrons with the accompaniment of the emission of
QED mesons with these spectral lines from their de-excitation.  By
judicial correlations with the probability of such collisions, those
QED meson spectral lines associated with the de-excitation of the
excited QED neutrons may be separated.  The on-set of these extra QED
meson lines arising from the decay of excited QED neutrons may be a
good signature of the on-set of deconfinement and a signature of the
quark-gluon plasma formation.

The search of the QED neutron may also be carried out by elastic and
inelastic collisions of the QED neutron with electrons in
semiconductors and superconductors, using detectors designed to search
for dark matter particles with a mass below 1 GeV/$c^2$. (See
Ref.\ \cite{Ess20} for a review and a list of references for such
detectors).  The proposed QED neutron has a predicted mass of 44.5
MeV.  It is a linear electric quadrupole and it interacts with an
electron by the electromagnetic quadrupole interaction $e^2a^2\cos^2
\theta/3|{\bb r}|^3$ where $\bb r$ is the radius vector from the
center of mass of the QED neutron to the electron, $\theta$ is the
opening angle between $\bb r$ and the QED neutron linear axis, and $a$
is the separation between the $u$ and the $d$ quark.  The separation
$a$ has been estimated in Table I to be 20.4 fm.  In a charged-coupled
device (CCD) detector, the collision between the QED neutron and the
electrons of the device may promote an electron from the valence band
to the conduction band.  In a transition-edge sensor (TES) detector,
which is a superconducting thin film held very close to its transition
temperature, the collision may deposit energy onto the thin film.  As
a result of the energy deposition, the superconducting film will make
a transition to the normal state signaled by an increase in its
resistance.  Another method makes use of a superconducting nanowire
single-photon detector (SNSPD) of a thin film of superconducting
material carrying a constant applied current.  The collision between
the QED neutron and the electrons of the detector may disrupt the
superconductivity and create a measurable signal.  Future studies of
the feasibility on the search for the QED neutron using semiconductor
and superconductor devices will be of great interest.

\section{Summary, Conclusions, and discussions}

The recent observations of the anomalous soft photons, the X17
particle, and the E38 particle have been consistently interpreted as
the production of QED mesons in \cite{Won10,Won11,Won14,Won20} arising
from the QED excitation of quarks and antiquarks in the QCD+QED gauge
field vacuum.  Heretofore our usual experience of quarks interacting
with gauge fields have been confined mainly to the situation with
quarks interacting with both QCD and QED gauge interactions
simultaneously, and they are invariably accompanied by gluon exchange
interactions.  How is it possible for a quark and an antiquark to
interact with the QED interaction alone without the QCD interaction?

We have presented arguments in the Introduction to show that a quark
and an antiquark can interact with the QED interaction alone if they
are produced in the range $(m_q + m_{\bar p}) < \sqrt{s} < m_\pi$.  We
note that the QED and QCD excitations of the quark vacuum can be
independent excitations.  The important ingredient in resolving the
conceptual difficulties rests on the fact that the quark currents and
gauge fields are not single-element quantities in QED and QCD
dynamics.  On the contrary, they are 3$\times 3$ matrices in color
space, forming a colors-singlet $\bb 1$ subgroup and a color-octet
$\bb 8$ subgroup.  The quark QCD and QED currents and gauge fields can
be independently excited in their respective subgroups.  Hence, there
can be QED excitations at the lower energies of many tens of MeV,
leading to the production of the QED mesons, with the QCD gauge fields
as non-participating background spectators.
 
If light quarks can interact with QED interactions in the
color-singlet sector with the QCD interaction as non-participating
background fields, a new frontier will be opened up for exploration
because quarks carry electric charges, electric charges of opposite
signs attract, and attractive interactions result in stable and
confined composite quark states.  There will be many composite systems
of light quarks and antiquarks in which the attractive QED forces
allow the composite system to form bound and confined states.  The
many quantum numbers that characterize the quarks will also add
complexities to the spectrum of these composite particles.  For
example, for light quarks with two flavors and $S$=0 moving in phase
and out of phases with each other, we show earlier that there can be
the $(I=0,I_3=0)$ state at 17.9$\pm$1.8 MeV and the $(I$=1,$I_3$=0)
state at 36.4$\pm$3.8 MeV.  We do not know whether QED mesons formed
by a light quark and a light antiquark of the same flavor are allowed
or not.  If they are allowed, then by using the semi-empirical formula
for the QED meson state energy developed in \cite{Won20}, one locates
theoretically the $S=0$ single-flavor excitation $d\bar d$ QED meson
state at 21.2$\pm$2.1 MeV and the $u\bar u$ QED meson state at
34.7$\pm$3.5 MeV.

The possible occurrence of the QED mesons can be tested by searching
for the decay products of two real photons, two virtual photons in the
form of double dilepton pairs, or a single dilepton pair.  The X17
particle observed in the decays of the $^4$He$^*$ and $^8$Be$^*$
\cite{Kra16,Kra19} with an $e^+e^-$ invariant mass of 17 MeV and the
state at 19$\pm$1 MeV in emulsion studies \cite{Jai07} match the
predicted mass of the isoscalar 0(0$^-)$ QED meson \cite{Won20}.  The
E38 MeV particle, observed in high-energy $p$C, $d$C, $d$Cu collisions
at Dubna \cite{Abr12,Abr19} with a $\gamma \gamma$ invariant mass of
about 38 MeV, matches the predicted mass of the isovector QED meson
\cite{Won20}.  These are encouraging experimental observations. The
QED $(u\bar u)$ and QED $(d\bar d)$ states have yet to be located and
identified experimentally.  Further experimental measurements in the
low invariant mass region will be of great interest.

The QED mesons are not the only color-singlet composite states arising
from quarks interacting in QED interactions in the color-singlet
subgroup, with the QCD gauge fields as non-participating spectators.
The QED neutron with two $d$ quarks and one $u$ quark with three
different colors can form a color-singlet composite system.  The QED
neutron can be stable because the attractive QED interactions between
two $d$ quarks and the $u$ quark overcome the weaker repulsion between
the two $d$ quarks.  With a phenomenological three-body model in 1+1
dimensions with an effective interaction between electric charges
extracted from Schwinger's exact QED solution, we find quantitatively
in a variational calculation that there is a QED neutron energy
minimum at a mass of 44.5 MeV.  The analogous QED proton with two $u$
quarks and a $d$ quark has been found to be too repulsive to be stable
and does not have a bound or continuum state.

Because of its composite nature, there will likely be many excited
states in the QED neutron.  The excited states are expected to decay
by emitting photons and/or QED mesons to make transitions to lower QED
neutron states.  One of the two $d$ quarks may decay into an $u$ quark
by way of the weak interaction.  However, because the QED proton does
not possess a stable bound state nor a continuum state of isolated
quarks, the rate of the QED neutron weak decay into a QED proton is
zero.

Among all QED neutron states, the ground QED neutron state located at
44.5 MeV distinguishes itself from higher excited QED neutron states
as a stable particle without decay products.  It can only decay by a
baryon-number non-conserving transition, which presumably has a very
long lifetime.  As a consequence, the lowest state QED neutron is a
dark neutron.  The QED antineutron ground states is likewise a dark
antineutron.  The only mode of destruction for a QED dark neutron and
a QED dark antineutron is their mutual annihilation, with the
production of photons and QED mesons.

It is of interest to discuss the relation between the QED neutron at
44.5 MeV and the QCD neutron at 939.6 MeV that is confined by QCD
gauge interactions.  We envisage that the QED neutron is an energy
minimum in the color-singlet subgroup of quark current and QED gauge
field, while the QCD neutron is an energy minimum in the color-octet
subgroup of quarks interacting with the exchanges of gluons between
quarks.  They are similar to the multiple energy minima which occur in
the collective energy surface of nuclear systems in
\cite{Hil53,Bra72}.  As members of different subgroups, they do not
mix in the lowest order but can be mixed in second-order perturbation
theory by photon interactions.  The mixing of QCD neutron in the QED
neutron may be small because the QED neutron can be connected to the
QCD neutron by electromagnetic interactions but the difference in
their longitudinal extensions of wave functions and the large energy
denominator in second-order perturbation theory may make the mixing
quite small.  The sea quarks arise predominantly from the splitting of
the gluons into quark-antiquark pairs which appear in the color-octet
subgroup of the product group.  Because the QED neutron resides in the
colors-singlet subgroup, the presence of the sea quarks
\cite{Gee10,Bed20} in the QCD color-octet subgroup will not likely
affect the stability of the QED neutron.

On account of their being predicted to be stable particles with a very
long lifetime without decay products, the QED dark neutrons and QED
dark antineutrons may be good candidate particles for a part of the
dark matter.  We envisage that in the early evolution of the Universe
after the big bang, the Universe will go through the quark-gluon
plasma phase with deconfined quarks and gluons.  As the primordial
matter expands and cools down the quark-gluon plasma undergoes a phase
transition from the deconfined phase to the confined phase, deconfined
quarks of three different colors may coalesce to form color-singlet
states.  While many of the color-singlet systems of three quarks have
sufficient energy to form hadrons, there may be some produced
color-singlet three-quark systems in which their total energy is below
the QCD neutron energy of about \break1 GeV.  These three-quark
systems may form QED neutron states that are bound by QED
interactions.  The de-excitation of the excited QED neutron state will
find its way down to the lowest energy state of the QED dark neutron.
Such QED dark neutrons and its excited states may occur at the
deconfinement-to-confinement phase transition of the quark-gluon
plasma and may be a signature of the deconfinement-to-confinement
transition of the quark gluon plasma in high-energy heavy-ion
collisions.  Self-gravitating assemblies of QED dark neutrons may be
stable astrophysical objects.  Because of its long lifetime,
self-gravitating QED dark neutron assemblies (and similarly QED dark
antineutron assemblies) of various sizes may be good candidates for a
part of the primordial dark matter produced during the
deconfinement-to-confinement phase transition of the quark gluon
plasma in the evolution of the early Universe.

In another matter, LIGO recently observed the merger of two neutron
stars through the detection of their gravitational waves in 2017
\cite{LIGO17}.  Such a merger will likely lead to the production of a
quark matter with deconfined quarks \cite{Bau19,Bau19a,Wei20}.  The
authors of \cite{Bau19,Bau19a} proposed a new signature for a
first-order hadron-quark phase transition in merging neutron stars,
which may provide the opportunity to study the properties of the
post-merger quark matter.  As the quark matter cools and undergoes the
deconfinement-to-confinement phase transition during the merging
process, the coalescence of deconfined quarks to become confined
quarks will produce QED neutrons in the post-merger environment.

In another astrophysical frontier, it has been suggested that
deconfined quark matter may be present in the core of a massive
neutron star \cite{Ann20}.  In such a neutron star, the transition
region close to the core may contain QED neutron matter arising from
the coalescence of deconfined quarks.  In matter of modeling massive
neutron star mergers of such a type, it may be necessary to take into
account the presence of such a QED neutron region.  Therefore, the
study of the equation of states and the the thermodynamical and phase
transition properties of the QED neutron matter will be of great
interest.  QED neutrons in the neutron star environment
will experience a strong magnetic field which will affect the QED
neutron in a significant way.  While the confinement of QED neutron
may continue to be operative, we expect that the additional magnetic
interaction will set the QED neutron into a rotational motion so that
the quarks in the QED neutron will go from a 1+1 dimensional dynamics
to a full three-dimensional dynamics.  It will be of interest to study
how the magnetic field will affect the QED neutron and its stability.

As it is suggested here that the confinement to deconfinement phase
transition at the early history of the Universe in the quark-gluon
plasma phase may generate the QED dark neutron assemblies as seeds for
the primordial dark matter, it will be of great interest to study
whether QED dark neutrons and/or its excited states may be produced in
high-energy heavy-ion collisions where quark gluon plasma may be
produced.  The detection of the QED dark neutrons may be made by
searching for the photons and/or QED mesons during the de-excitation
from its excited states.  The de-excitation of the excited QED neutron
states will yield photons and QED mesons with its own characteristic
QED meson emission spectrum.  The capability of precise dilepton
measurements in high-energy heavy-ion collisions
may make it possible to study the spectrum of the produced QED mesons
through their decays into two virtual photons, as discussed in
Appendix F.  We may rely on the presence of these emitted photons or
produced QED mesons to reconstruct the spectrum of the QED neutron.

For simplicity in the present first survey of the QED neutron, we have
neglected the spin degree of freedom.  While the spin will not likely
affect the stability, the quark confinement, and the gross
structure of the QED neutron, it will play a significant role in the
fine structure and the spectrum of the QED neutron.  The spin degree
of freedom, along with the orbital angular momentum, the collective
rotation, and the collective vibration should be taken into account in
future studies.  Theoretical investigations on the internal structure
and the energy spectrum of the QED neutron will be valuable to assist
the detection of the produced QED neutrons.  How the QED mesons and
QED neutron interact with themselves and with hadrons will open up
another avenue to explore the interplay between species from the same
branch and from different branches of the quark family of the Standard
Model.  Furthermore, the possibility of the many-body interaction
between QED dark neutrons forming a bound multi-QED-neutron system and
the interaction between the QED neutron matter and other standard QCD
matter will add other dimensions to the complexity of matter
associated with the QED dark neutron.  It will be of great interest to
extend the frontier of QED neutrons both theoretically and
experimentally.

\section{Acknowledgments}

The author would like to thank Profs. H. Georgi, Y. Jack Ng, Lai-Him
Chan, A. Koshelkin, Gang Wang, H. Sazdjian, G. Wilk, Pisin Chen, Larry
Zamick, Jia-Chao Wang, Scott Willenbrock, and J. Stone for helpful
communications and discussions.  The research was supported in part by
the Division of Nuclear Physics, U.S. Department of Energy under
Contract DE-AC05-00OR22725.

\begin{appendix}

\section{Schwinger's boson and  massless  fermions 
in 1+1 dimensional QED}

Our goal is to study the stability of color-singlet states involving
three light quarks in QED interactions.  To pave the way for such an
investigation, we would like to sharpen our theoretical tools by
examining the analogous two-body problem of a massless
fermion-antifermion pair in QED, for which the exact solution from
Schwinger is already known \cite{Sch62,Sch63}.

Schwinger showed previously that in 1+1 dimensions, massless fermions
and antifermions interacting with the QED gauge interaction with a
coupling constant $e$=$e_\2d$$\equiv$$g^\qedu_\2d$ give rise to a
bound boson with a mass $m$, given by \cite{Sch62,Sch63}
\begin{eqnarray}
m=\frac{e} {\sqrt{\pi}},
\label{2p1}
\end{eqnarray}    
where the QED coupling constant $e$ has the dimension of a mass in 1+1
dimensions.  In terms of the description of $e$ as a unit of charge
and $Q$ as the charge number, the fermion can be described as
possessing a charge $e$ and a charge number $Q=1$ and the antifermion
a charge $(-e)$ and a charge number $Q=(-1)$ in the Schwinger model.

A derivation of Schwinger's exact solution of (\ref{2p1}) can be found
in \cite{Sch63} and explained in details in Chapter 6 of \cite{Won94}.
Recent generalizations and extensions of the Schwinger model have been
presented in \cite{Geo19,Geo19a,Geo20}.  It is illuminating to review
its salient points here to see in what way we may treat Schwinger's
boson as a relativistic two-body problem.  Schwinger's exact solution
can be obtained self-consistently as a many-body field theory problem
involving the response of all fermions in the presence of a perturbing
gauge field $A^\mu$.  We start by considering a vacuum state in which
all the negative-energy states of the massless fermions in the Dirac
sea are occupied.  A disturbance in the fermion density and/or fermion
current $j^\mu$ will generate a perturbing gauge field $A^\mu$.  The
presence of the perturbing $A^\mu$ induces a change of the gauge
phases of fermion field operator $\psi$ through the Dirac equation
\begin{eqnarray}
\gamma_\mu(i\partial_\mu-eA^{\mu})\psi(x)=0,
\label{a2}
\end{eqnarray}
where $\gamma^\mu$ are the gamma matrices in 1+1 dimensions.  The
change of the gauge phases of fermion field operator $\psi$ in turn
lead to a change of the fermion current $j^\mu$.  By imposing the
Schwinger modification factor, $e^{ie\int_{x'}^x A_\mu
  (\xi)d\xi^\mu}$\!\!, to ensure the gauge invariance of the fermion
Green's function, the induced fermion current $j^\mu$ is given
implicitly as a function of the perturbing gauge field $A^\mu$ by
\begin{eqnarray}
j^\mu (x) && = {-e\over 2} \biggl \lbrace \lim_{{x^0 = {x^0}'} \atop
  {x^1={x^1}'-\epsilon}} \!\!+\!\! \lim_{{x^0 = {x^0}'} \atop
  {x^1={x^1}'+\epsilon}} \biggr \rbrace\ tr \biggl [e^{ie\int_{x'}^x
    A_\mu (\xi)d\xi^\mu} \nonumber\\ && \hspace*{2.8cm}\times \langle
  T({\bar \psi}(x')\gamma^\mu \psi(x)\rangle \biggr ],
  \label{a3}
\end{eqnarray} 
where $T$ is the time-order operator.  Upon evaluating the above
limits from the left and from the right at the space-time point
$x=(x^0,x^1)$, the fermion current singularities from the left and
from the right cancel and the induced gauge-invariant fermion current
$ j^\mu$ is found to be related explicitly to the perturbing gauge
field $A^\mu$ by
\begin{eqnarray}
j^\mu = -\frac{e}{\pi}\left [ A^\mu - \partial ^\mu
  \frac{1}{\partial^\lambda \partial _\lambda }\partial \nu A^\nu
  \right ] .
\label{a4}
\end{eqnarray}
The resultant fermion current $j^\mu$ in turn leads to a new gauge
field ${\tilde A}^\mu$, through the Maxwell equation,
\begin{eqnarray}
\partial_\nu F^{\mu \nu} = \partial_\nu ( \partial ^\mu {\tilde A}^\nu
- \partial^\nu {\tilde A}^\mu) = e j^\mu .
\label{a5}
\end{eqnarray}
From Eqs.\ (\ref{a4}) and (\ref{a5}), the self-consistency of the
resultant gauge field ${\tilde A}^\mu$ matching the initial perturbing
gauge field $A^\mu$ leads to the gauge field satisfying the
Klein-Gordon equation 
\begin{eqnarray}
-\square A^\mu -\frac{e^2}{\pi} A^\mu =0.
\end{eqnarray}
Therefore, massless fermions and antifermions interacting
non-perturbatively in a gauge field with a coupling constant $e$ in
1+1 dimensions result in a boson field with a quanta of mass
$m$=$e/\sqrt{\pi}$.

It is clear from the above review that the exact solution of the boson
state does not lend itself readily to a simple quantum mechanical
two-body problem of valence fermion (quark) and valence antifermion
(antiquark) involving a simple fundamental two-body interaction,
because it involves concepts of massless fermions, gauge invariance,
gauge field self-consistency, and the cancellation of the fermion
current singularities that are beyond the conventional two-body
problems with a simple two-body interaction.  In spite of this being
the case, it is desirable to construct a phenomenological two-body
model for the valence fermion and antifermion with an effective
interaction $\Phi$ that can be calibrated to contain the basic
properties of the theory and to yield Schwinger's exact result.
Examples of such an approach can be found in the successes of
relativistic and non-relativistic hadron spectroscopy where the
non-perturbative QCD solution involving the lattice gauge theory is
approximated by a two-body theory with phenomenological effective
interactions (see for example,
\cite{Ruj75,God85,Bar92,Won00,Won01,Cra83,Cra92,Cra06,Cra07,Cra09,Won01a}).
Being a phenomenological two-body theory, such a theory and its
generalizations will need to be worked out with considerable
theoretical support and persistent confrontation with experiments so
that it can be refined and readjusted, should new experimental data
and new theoretical predictions become available.  The present
investigation on the stability of the QED neutron represents an
exploration along such lines.

An additional advantage of a successful phenomenological two-body
problem treatment rests on it ability to simplify the calculations, to
retain the essential features, to provide an intuitive understanding,
and to help solve problems that may not be solvable in a full
treatment of the field theory, paving the way for our analysis on the
stability of the three-quark system in section 2.

\section{The separation and the independence  of the color-singlet
  and color-octet $q\bar q$ excitations }

It is instructive to review how the color-singlet and color-octet
$q\bar q$ 
excitations can arise in a quark-QCD-QED system.  We start with the
quark-QCD-QED vacuum which is the lowest energy state with quarks
filling up the negative-energy Dirac sea and interacting in QCD and
QED interactions.  We introduce a disturbance which creates one or
many $q\bar q$ pairs, as for example, (i) during the de-excitation of
a highly-excited nuclear state with a proton pulling outside an alpha
particle core, (ii) in a high-energy nuclear collision, or (iii) in an
excited system of a quark and an antiquark stretching out from each
other at high energies after the annihilation of a high energy
$e^+e^-$ pair, as represented schematically by Figs.\ (1a), (1b), and
(1c) of \cite{Won22b}.  Final state interactions allow the
creation of only those final $q\bar q$ states at eigenenergies of the QCD
or QED mesons because the density of final states away from the meson
eigenenergies is zero on account of the confinement of quarks.
 
We focus our attention on one of the lowest energy produced $q\bar q$
pairs.  For the created $q\bar q$ pair to be in a QCD or QED meson
state, there must be a direction of dominant motion of the quark and
the antiquark which can be taken to be the longitudinal direction.  We
can infer the occurrence of transverse confinement of the created
$q\bar q$ pair and the created QCD and QED gauge fields at the moment
of the their creation, from Polyakov's results of the confinement of
opposite charges in compact Abelian and non-Abelian gauge theories in
(2+1)D \cite{Pol77,Pol87}.  The cylindrical flux tube arising from the
subsequent longitudinal stretching of the initial quark and antiquark
will remain transversely confined, as discussed in
\cite{Won11,Won22b}.  We can thus study the longitudinal dynamics of
the quark and the antiquark system by idealizing the cylindrical flux
tube as a one-dimensional string in (1+1)D, with its information on
the transverse flux tube radius $R_T$ stored in the new gauge field
coupling constant $g$=$g_\2d$=$g_\4d/\sqrt{\pi}R_T$ in (1+1)D, and its
transverse quark mass $m_T$ obtained from the eigenvalue of a
transverse equation of motion of the quark in the flux tube
\cite{Won10,Won20,Kos21}.  Because the masses of light quarks are
small, it is reasonable to assume light quarks to be massless so that
Schwinger's model of massless charges can be applied.

We are now in a position to examine the color degrees of freedom in
the longitudinal dynamics of quarks and antiquarks in a general
quark-QCD-QED system in the idealized (1+1)D.  As we mentioned in the
Introduction, the quark current $j^\mu$ and the gauge fields $A^\mu$
are not single-element functions.  They are in fact 3$\times 3$ color
matrices.  Quarks in color-triplet $\bb 3$ representation and
antiquarks in color $\bb 3^*$ representation form the product group of
$\bb 3$ $\otimes$ $\bb 3^*$=$\bb 1$ $\oplus$ $\bb 8$, which contains
the color-singlet $\bb 1$ subgroup and the color-octet $\bb 8$
subgroup of generators.

In the general case, the color-octet QCD gauge fields $A^\mu(x)$
=$\sum_{i=1}^8 A^\mu_ i (x)t^i$ contains eight non-Abelian degrees of
freedom and their couplings will lead to color excitations, the
majority of which will not lead to stable collective excitations.  To
get stable collective excitations, we represent the color-octet
degrees of freedom by a single unit vector $\tau^1$ oriented randomly
in the eight-dimensional SU(3) color generator space
\cite{Won10,Won20,Kos21},
\begin{eqnarray}\label{cosines}
 \tau^1=\sum_{i=1}^8 n_a t^a, ~~~\text{with}~~~\sqrt{n_1^1+n_2^2+...+n_8^2}=1,
\end{eqnarray}
where $n_a=2{\rm Tr }\{\tau^1 t^a\}$, and we restrict the dynamics of
the QCD gauge fields only to the gauge field amplitude $A^\mu_1(x)$
along the $\tau^1$ direction, while keeping the $\tau^1$ orientation
fixed.  The gauge fields of the quark-QCD-QED system are then
described by the color-singlet amplitudes $A^\mu_0(x)$ for QED
dynamics and the color-octet amplitude $A^\mu_1(x)$ for QCD dynamics
as \cite{Won10,Won20,Kos21}
\begin{eqnarray}
A^\mu (x) =  A^\mu_0 (x) \tau^0 + A^\mu _1 (x)\tau^1=\sum_{\lambda=0}^1
A_\lambda^\mu(x) \tau^\lambda,
\label{b8}
\end{eqnarray}
where $\tau^0=t^0$, $2{\rm Tr}(\tau^{\lambda} \tau^{\lambda'})
=\delta^{\lambda\lambda'} $, and $\lambda,\lambda'=0,1$.  The quark
currents $j^\mu (x)$ can be likewise represented by the color-singlet
and color-octet current components as
\begin{eqnarray}
j^\mu (x) =  j^\mu_0 (x) \tau^0 + j^\mu _1 (x)\tau^1=\sum_{\lambda=0}^1
j_\lambda^\mu(x) \tau^\lambda.
\label{b9}
\end{eqnarray}
Because $\tau^0$ and $\tau^1$ commute, the gauge fields in this
restricted subspace are Abelian.

We can now examine the dynamics of the quarks fields and the gauge fields.
We start with initial gauge fields $A_\lambda^\mu$ which affect the
quark field.  The Dirac equation (\ref{a2}) for the quark field is
\begin{eqnarray}
\gamma_\mu \left (i\partial^\mu+\sum_{\lambda=0 }^1 g^\lambda A_\lambda^\mu (x)  \tau^\lambda \right ) \psi(x)=0.
\end{eqnarray}
After obtaining the solutions to the above Dirac equation for the
quark field $\psi(x)$, we evaluate the quark current $j^\mu_\lambda (x) $ as
a function of the applied inital gauge field $A^\mu_\lambda(x) $,
\begin{eqnarray}
&&j_\lambda^\mu (x)  = {1\over 2} \biggl \lbrace \lim_{{x^0 = {x^0}'} \atop
  {x^1={x^1}'-\epsilon}} \!\!+\!\! \lim_{{x^0 = {x^0}'} \atop
  {x^1={x^1}'+\epsilon}} \biggr \rbrace
  \label{nb11}
  \\
  &&~~~tr \biggl [e^{i\int_{x'}^x
    \sum_{\lambda'  }^1 (-g^{\lambda' })  A_{\lambda'}^\mu (\xi)\tau^{\lambda'}
     d\xi_\mu} 
     \langle
  T({\bar \psi}(x')\gamma^\mu \tau^\lambda \psi(x)) \rangle \biggr ].
  \nonumber
\end{eqnarray}
The induced quark current $j_\lambda^\mu(x)$ as a function of the
initial applied gauge field $A^\mu_\lambda$ is found to be
\cite{Kos21}
\begin{eqnarray}
j_\lambda ^\mu (x)=  \frac{g^{\lambda }}{\pi}\left ( A_{\lambda }^\mu(x)  - \frac{\partial _\nu \partial ^\mu}{\partial^2} A^\nu_{\lambda } (x) \right ),
\label{b12}
\end{eqnarray}
where $\lambda$=0 for QED, $\lambda$=1 for QCD.  We note specially that in the
above Eq.\ (\ref{b12}), the induced current $j_\lambda^\mu(x)$ of type
$\lambda$ depends only on the initial gauge field $A_{\lambda
}^\mu(x)$ of same type $\lambda$ and does not depend on the gauge
field of the other type, even though both types of gauge fields are present in the sum over $\lambda'$ in the Schwinger gauge-invariant  exponential factor.
 Such a dependence (and independence) arises because in
obtaining the above result in Eq.\ (\ref{b12}), we expand the Schwinger gauge-invariant exponential factor  in Eq.\ (\ref{nb11}), and take the trace over
the color space involving the $\tau^{\lambda'}$ generator from the
Schwinger factor, and the $\tau^\lambda$
generator from $T\langle \bar \psi \gamma^\mu \tau^\lambda \psi
\rangle$.  Because of the orthogonality of the
color generators,
2tr$\{\tau^{\lambda'}\tau^{\lambda}\}$=$\delta^{\lambda' \lambda}$,
the contribution from the interaction of the other type is zero
\cite{Kos21}.  In physical terms, the above mathematical result means
that to be observable a color-singlet quark density oscillation must
remain colorless\footnote{The physical quark density is the time-component of the quark current, $j_\lambda^0(x)$, and its temporal dependence gives the quark density oscillation.  In mathematical terms, the observability of the quark density oscillation corresponds to a non-zero value of the quark current upon 
taking the trace over the color space.}
 and cannot become colored, when the color-singlet quark density oscillation  absorbs a gauge
boson.  Hence, it can only absorb a QED gauge boson $A^\mu_0$ but not
a QCD gauge boson $A^\mu_1$ to make the color-singlet quark density
oscillation observable.  On the other hand, a color-octet quark
density $j_1^\mu(x)$ oscillation cannot become observable until it bleaches its
octet color to become colorless.  Hence it can only absorb a QCD gauge
boson $A^\mu_1$ but not a QED gauge boson $A^\mu_0$ to make the
color-octet quark density $j_1^\mu(x)$ oscillation observable.  Therefore, in the
collective dynamics of the quark-QCD-QED medium, quark currents of the
type $\lambda$, $j_\lambda^\mu(x)$,  is affected only by gauge fields of the same type
$\lambda$, ,$A_\lambda^\mu(x)$, as shown in Eq.\ (\ref{b12}).
 
The induced quark current $j_\lambda^\mu(x)$ generates a new gauge
fields $\tilde A ^\mu_\lambda(x) $ through the Maxwell equation, which
in the Abelian approximation of Eq.\ (\ref{b8}) is
\begin{eqnarray}\label{Maxwell-1}
-\partial_\nu \partial^\nu \tilde A_{\lambda }^\mu(x) +    \partial_\nu \partial^\mu \tilde A_{\lambda }^\nu (x) = 
g^{\lambda } j_{\lambda }^\mu(x) .
\label{b13}
\end{eqnarray}
Stable self-consistent collective dynamics of the quark field and the
gauge fields can be obtained when the newly generated gauge fields
$\tilde A^\mu_\lambda$ are the same as the initial applied gauge
fields $A^\mu_\lambda$.  Such self-consistency can be achieved by
setting $\tilde A^\mu_\lambda=A^\mu_\lambda$ and substituting
Eq.\ (\ref{b13}) into Eq.\ (\ref{b12}).  We get the Klein-Gordon
equation for the currents
\begin{eqnarray}
&&-\partial^\nu \partial_\nu j_{\lambda } ^\mu   = m_{\lambda}^2 j_{\lambda } ^\mu , 
\end{eqnarray}
which corresponds to the occurrence of a boson of a stable and
independent collective excitation of the quark-QCD-QED medium, with a
mass $m_\lambda$ given by
\begin{eqnarray}\label{mass_a}
&&m_{\lambda}^2 = \frac{( g^{\lambda })^2 }{\pi},~~
\begin{cases}
  \lambda=0 & \text{for QED} \cr
  \lambda=1 &\text{for QCD} \cr 
\end{cases}.
\label{b15}
\end{eqnarray}
From another perspective, the separation and the independence of the color-singlet and the
color-octet excitations are possible because the gauge-invariant
relations between the charge currents $j^\mu$ and the gauge fields
$A^\mu$ in (1+1)D in Eqs.\ (\ref{a4}) and (\ref{a5}) (and similarly in
Eqs. (\ref{b12}) and (\ref{b13})) are each a linear function of
$j^\mu$ and $A^\mu$.  As a consequence, there is a principle of
superposition of currents and gauge fields of different color
components in color-space.  Thus, the quark-QCD-QED medium possesses
stable and independent collective QCD and QED excitations with
different bound states masses $m_\lambda$, depending on the coupling
constants $g^\lambda $.  From Eq.\ (\ref{b15}), the ratio of the QED
meson mass to the QCD meson mass is of order
\begin{eqnarray}
\frac{m_{\qedu\,{\rm meson}}}{m_{\qcdu\,{\rm
    meson}} }
    \sim \frac{g^\qedu}{g^\qcdu} 
    \sim \!\sqrt{\frac{\alpha_{{}_\qedd}}{\alpha_{{}_\qcdd}}}
    \sim \!\sqrt{\frac{1/137 }{0.6 }}
     \sim \frac{1}{9},
\label{b16}
\end{eqnarray}
which is Eq.\ (1).
 
\section{Relativistic  two-body problem}

The relativistic two-body wave equations for the wave function $\Psi$
for QED interactions in 1+1 dimensions consist of two mass-shell
constraints on each of the interacting particles
\cite{Cra83,Cra92,Cra06,Cra07,Cra09,Cra10,Won01a,Saz87,Saz89},
\begin{subequations} 
\begin{eqnarray}
  {\cal H}_{1}|\Psi\rangle =\biggl  \{ p_{1}^{2}-m_{1}^{2}-\Phi _{12}(x_{12})\biggr \} |\Psi\rangle =0,
\label{3p1a}
\\
{\cal H}_{2}|\Psi\rangle =\biggl  \{  p_{2}^{2}-m_{2}^{2}-\Phi _{21}(x_{21}) \biggr \} |\Psi\rangle=0.
\label{3p1b}
\end{eqnarray}
\end{subequations} 
As emphasized in Appendix A, Schwinger's exact solution in field
theory resulting in longitudinal confinement is in fact the solution
of the collective motion for a many-body problem of great complexity.
It is not a simple two-body problem of a fundamental interaction.  In
transcribing the Schwinger field theory problem as a phenomenological
two-body problem, there are choices of different forms of the two-body
effective interaction.  An alternative form uses the QED confining
linear interaction in the Coulomb gauge as the time-like vector
potential $V_{12}(r)$ in a minimum substitution form, and it leads to
$[(p_i^0- V_{12}(x_{12}))^2 - {\bb p}_i^2-m_0^2]\Psi=0$.  However,
such an effective interaction has the unpleasant feature that the
$V_{12}^2$ term in the above Schr\"odinger-type equation leads to
solutions whose behavior do not match the behavior of the Schwinger
solution at large separations.  It becomes necessary to introduce an
additional scalar confining interaction $S_{12}(x_{12})$ so that the
wave equation becomes $[(p_i^0- V_{12}(x_{12}))^2 - {\bb
    p}_i^2-(m_0^2+S_{12}(x_{12}))^2]\Psi=0$.  The choice of
$V_{12}=S_{12}$ used in \cite{Cra10} leads to a cancellation of the
$V_{12}^2$ term with the $S_{12}^2$ term at large separations and a total effective
interaction similar to the $\Phi$ used here.  In view of the
phenomenological nature of the confining interaction, the present
effective confining interaction $\Phi_{12}$ in Eq.\ (\ref{3p1b})
serves well as a description that can match the confinement property
of the Schwinger solution.

We would like to calibrate the above effective interaction $\Phi_{ij}$
by comparing the solution of the above two-body problem with
Schwinger's exact QED solution in 1+1 dimensions.  We construct the
total Hamiltonian ${\cal H}$ from these constraints by
\begin{equation}
{\cal H}=\sum_{i=1}^2 {\cal H}_{i}.
\end{equation}
In order that each of these constraints be conserved in time we must
have
\begin{equation}
\lbrack {\cal H}_{i},{\cal H}]|\psi \rangle=i\frac{d{\cal H}_{i}}{d\tau }
|\Psi \rangle=0.
\end{equation}
As a consequence, the above equation leads to the compatibility
condition between the two constraints
\cite{Cra83,Cra92,Cra06,Cra07,Cra09,Cra10,Won01a},
\begin{equation} 
\lbrack {\cal H}_{i},{\cal H}_{j}]|\Psi \rangle=0.  \label{3}
\end{equation}
Since the masses $m_1$ and $m_2$  commute with the operators, this implies 
\begin{eqnarray}
&&\biggl  (  [ p_{1}^{2}\Phi _{21}(x_{21})]    -   [ p_{2}^{2}\Phi _{12}(x_{12})]
\biggr  )|\Psi \rangle=0.
\label{3p5}
\end{eqnarray}
The above equation cannot be satisfied if $\Phi_{12}(x_{12})$ $\ne$
$\Phi_{21}(x_{21})$.  The simplest way to satisfy the above equation
is to take
\begin{eqnarray}
\Phi _{12}(x_{12})=\Phi _{21}(x_{21})=\Phi(x_{12}), 
\end{eqnarray}
which is the relativistic analogue of Newton's third law.  The
compatibility condition (\ref{3p5}) then requires the effective
interaction $\Phi(x_{12})$ to depend only on the coordinate
$x_{12\perp}$ transverse to the total momentum $P$=$p_1+p_2$,
\begin{eqnarray}
\Phi (x_{12})=\Phi (x_{12\perp}),
\end{eqnarray}
\begin{eqnarray}
\text{where} ~~ x_{12 \perp} = (x_1-x_2) - \frac{(x_1-x_2)\cdot P}{P^2}P~.
\end{eqnarray}
We shall work in the CM system where the total momentum $P$ is
$(P^0,P^1)$=$(M,0)$, $M$ is the invariant mass of the composite
system, and the relative coordinate $x_{ij\perp}$=$(x_1-x_2)$ involves
only spatial coordinates $x_1$ and $x_2$.  The particle momentum $p_i$
can be separated out into a component $\epsilon_i$ parallel to $P$ and
a component $q_i$ transverse to $P$ as
\begin{eqnarray}
p_i &&=(\epsilon_i, q_i)=\epsilon_i \frac{P}{\sqrt{P^2}}+q_i , ~~~i=1,2 ,~~~~\\
\text{where} \hspace*{0.5cm}\epsilon_i&&=\frac{p_i\cdot P}{\sqrt{P^2}},
~~~~~~ \text{and}~~~~~~  q_1+q_2=0.
\end{eqnarray}
In terms of $\epsilon_i$, the invariant mass of the composite system
$M$ is given by
 \begin{eqnarray}
M=P^0=\epsilon_1+\epsilon_2.
\end{eqnarray}
The two-body wave equations (\ref{3p1a}) and (\ref{3p1b}) in the CM
system becomes
\begin{subequations}
\label{eq18}
\begin{eqnarray}
 \epsilon_1^2 | \Psi  \rangle =  \left \{  q_1^2 +m_1^2 + \Phi (x_{12\perp})  \right \} |\Psi \rangle ,
\label{3p13a}
\\
 \epsilon_2^2 | \Psi  \rangle =  \left \{  q_2^2 +m_2^2 +  \Phi(x_{12\perp})  \right \} |\Psi \rangle. \label{3p13b}
\end{eqnarray}
\end{subequations}
Because  $q_2=(- q_1)$, the second equation of the above is simply 
\begin{eqnarray}
(\epsilon_2^2 - \epsilon_1^2)  <| \Psi  \rangle= (m_1^2-m_2^2) | \Psi  \rangle.
\label{3p14}
\end{eqnarray}
It is only necessary to solve for the eigenstate of the first
Schr\"odinger-type equation (\ref{3p13a}) to obtain $\epsilon_1$, and
the quantity $\epsilon_2$ can be obtained as an algebraic equation
from (\ref{3p14}).  The knowledge of $\epsilon_1$ and $\epsilon_2$
(taken to be positive) then gives the invariant mass $M$ of the
interacting two-body system.

\section{Schwinger's QED boson as a relativistic two-body problem
  in 1+1 dimensions}

The brief summary presented in Appendix A makes it plain that the
Schwinger boson that is confined and bound in 1+1 dimensions is in
fact a non-linear self-consistent solution of a many-body system of
great complexity.  It is desirable to construct a phenomenological
two-body problem for QED in 1+1 dimensions involving a valence fermion
and a valence antifermion with an effective phenomenological
interaction $\Phi(x_{12 \perp})$ that can be calibrated to contain the
basic properties of the theory and to yield Schwinger's exact QED
result in 1+1 dimensions.
 
Accordingly, we consider the two-body wave equations (\ref{3p13a}) and
(\ref{3p13b}) (or (\ref{3p14})) in the CM system for two charge
particles with charge numbers $Q_1$ and $Q_2$, interacting with a
phenomenological effective QED interaction $\Phi(x_{12\perp})$
\begin{eqnarray}
\Phi(x_{12\perp} )=\frac{2\epsilon_1 \epsilon_2}{\epsilon_1+\epsilon_2}(- Q_1 Q_2) \kappa | x_1-  x_2|.
\label{4p1}
\end{eqnarray}
The effective QED interaction $\Phi(x_{12\perp})$ in 1+1 dimensions
has been chosen such that:

\begin{enumerate}

\item  
In the Coulomb gauge, the interaction energy between charges $Q_1e $
and $Q_2e $ in 1+1 dimensional QED is $(-Q_1 Q_2 e^2/2) | x_1 - x_2|$
\cite{Col76}, which is indeed confining for the attractive interaction
between unlike charges.  It is reasonable to use such a spatially
linear interaction in the phenomenological two-body problem.  The
value of the coefficient parameter $\kappa$ of such a linear potential
in (\ref{4p1}) will be affected by the use of the phenomenological
reduced mass factor, Schwinger's self-consistency condition on the
gauge field, and the gauge invariance constraint.  It is therefore
appropriate to extract $\kappa$ phenomenological from Schwinger's
exact solution.

\item
The quantity $\kappa$ is proportional to the square of the coupling
constant $e^2$.  The exact value of $\kappa$ will be chosen to give
Schwinger's solution of $m=e/\sqrt{\pi}$ for a massless
fermion-antifermion pair interacting in the QED interaction.

\item
The effective interaction contains the charge factor $(-Q_1 Q_2)$,
which leads to an attractive interaction if $Q_1 Q_2 <0$, and a
repulsive interaction if $Q_1 Q_2 >0$, as in standard QED interaction
in quantum electrodynamics.

\item
The reduced mass factor $2(\epsilon_1 \epsilon_2)/(\epsilon_1 +
\epsilon_2)$ has been chosen to give the proper reduced mass in the
non-relativistic two-body wave equation.  We have however used the
particle energy $\epsilon_i$ in lieu of the rest mass $m_i$, to make
it applicable also to the massless limit.  The reduced mass factor
depends on the eigenvalues $\epsilon_i$ which should be
self-consistently determined by the wave equations (\ref{3p13a}) and
(\ref{3p13b}).

\end{enumerate}

\begin{figure} [h]
\centering 
\includegraphics[scale=0.40]{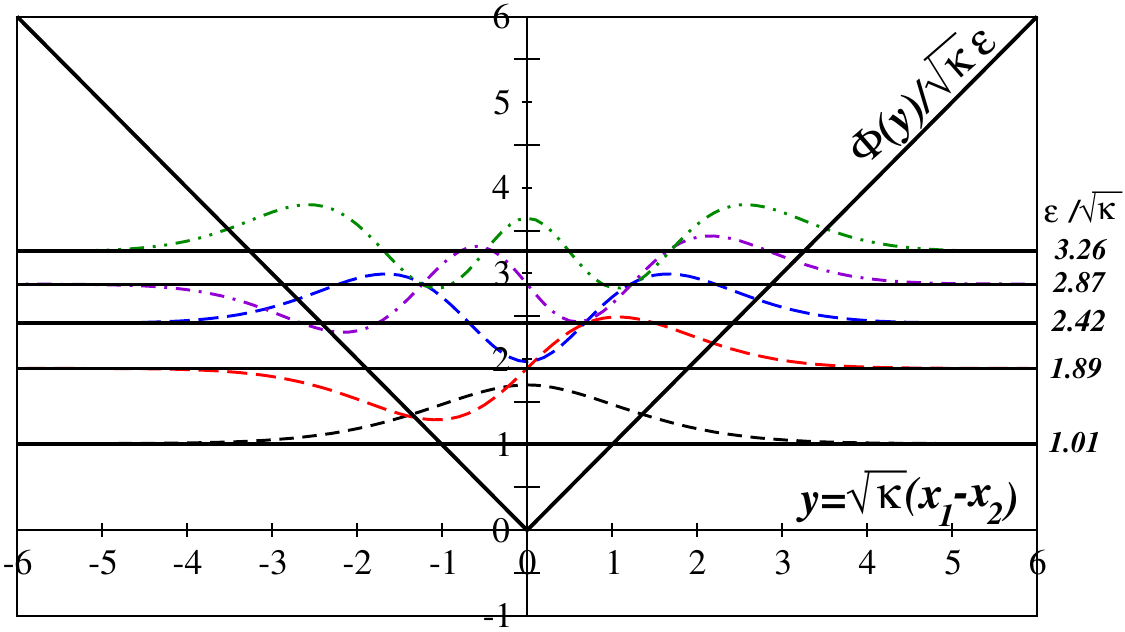}
\caption{ The solid curve gives the phenomenological effective
  potential $\Phi(y)/\sqrt{\kappa}\epsilon$ for a fermion and an
  antifermion interacting in the QED two-body problem in 1+1
  dimensions.  The energies of the lowest five eigenstates are
  indicated by horizontal lines with their corresponding wave
  functions $\Psi(y)$ as curves on these lines.  The parameter
  $\kappa$ in the phenomenological effective potential
  $\Phi(y)/\sqrt{\kappa}\epsilon$ is obtained by matching the solution
  for the lowest-energy state of the two-body problem with Schwinger's
  exact solution.  }
  \label{linV}
\end{figure}

Our present task is to obtain the eigenstates for the wave equation
(\ref{3p13a}) with the effective potential of (\ref{4p1}). 
Schwinger's case of massless fermion and antifermion corresponds to
$Q_1$=1, $Q_2=-1$, $m_1=m_2=0$, $\epsilon_1=\epsilon_2=\epsilon$, and
$M=2\epsilon$.  Using the dimensionless variable
\begin{eqnarray}
 y=\sqrt{\kappa}(  x_1 - x_2) ,
\end{eqnarray}
the effective interaction $\Phi(y)/\kappa$ is then
\begin{eqnarray}
\frac {\Phi(y)}{\kappa}=     \frac{\epsilon}{\sqrt{\kappa}}|y|,
\label{4p3}
\end{eqnarray}
and the wave equation (\ref{3p13a}) becomes
\begin{eqnarray}
&&\left\{ -\frac{\partial^2}{\partial y^2}+\frac{\epsilon}{\sqrt{\kappa}} |y| -  \frac{\epsilon^2}{\kappa}\right\} \Psi (y)=0.
\label{4p4}
\end{eqnarray}
We show in Fig.\ \ref{linV} the dimensionless effective potential,
$\Phi(y)/\sqrt{\kappa}\epsilon$, expressed in units of $\sqrt{\kappa}
\epsilon$.  It is a linearly-rising function of the dimensionless
spatial separation $|y|$ between the charges, expressed units of
$1/\sqrt{\kappa}$.  The solution of the wave equation with a linearly
rising interaction is the Airy function.  The wave equation
(\ref{4p4}) becomes the Airy equation
\begin{eqnarray}
&&\left\{ \frac{\partial^2}{\partial z^2}- \biggl ( |z|
-(\frac{\epsilon}{\sqrt{\kappa}})^{4/3}
 \biggr ) \right\}  \Psi(z)=0, 
\\
&&\text{where}  ~~z=(\frac{\epsilon }{\sqrt{\kappa}})^{1/3} |y| .
\end{eqnarray}
The solution satisfying the boundary condition at large $|y|$ is
\begin{eqnarray}
\Psi (z) = {\rm Ai}(|z| - (\frac{\epsilon}{\sqrt{\kappa}})^{4/3}).
\label{4p6}
\end{eqnarray}
The eigenstate is obtained by matching the wave function and its
derivative at $z$=0.  There are two types of eigenstates with even
and odd parities:
\begin{eqnarray}
&&\text{even parity:} ~~~\Psi '(z)|_{z=0} ={\rm Ai}'(- (\frac{\epsilon}{\sqrt{\kappa}})^{4/3})=0,
\\
&&\text{odd parity:}~~~~\Psi (z)|_{z=0} ={\rm Ai}(- (\frac{\epsilon}{\sqrt{\kappa}})^{4/3})=0.
\end{eqnarray}
We label the locations where the wave function or its derivative 
are zero as $(-a_s)$ with ${\rm Ai}(-a_s)=0$ or\break  ${\rm Ai}'(-a_s)$=0.
Then the eigenvalues of the wave equations $\epsilon$ are given by 
\begin{eqnarray}
(\frac{\epsilon}{\sqrt{\kappa}})^{4/3}=a_s, ~~~ \text{or}~~~ \epsilon=a_s^{3/4} \sqrt{\kappa}.
\end{eqnarray}

\begin{table}[H]
\centering
\caption{Solution of the two-body problem with the effective
  interaction $\Phi(y)/\sqrt{\kappa}\epsilon$ for the lowest states.
  Here $n$ is the number of nodes of the two-body wave function, and
  $M/\sqrt{\kappa}$ is the dimensionless measure of the composite
  particle mass.}
  \label{tab2}
\vspace*{0.3cm}
\begin{tabular}{|c| c| c |c|c|c|c|}
\hline
$n$& Parity &\!$a_s$\!&$\epsilon/\sqrt{\kappa}$  &  $M/\sqrt{\kappa}$  
&\!$\sqrt{\langle y^2\rangle }$ \!&\!$\sqrt{\langle q_1^2\rangle }/\sqrt{\kappa}$\!\\
\hline
 0 & even & 1.02& 1.01  & 2.02   & 0.862 & 0.60\\
1 & odd & 2.34 & 1.89  & 3.98   & 1.38  &  1.09\\
2 & even & 3.25 &2.42  & 4.84   & 1.78 & 1.41\\
3 & odd & 4.09  & 2.87 & 5.74   & 2.10 & 1.66\\
4 & even & 4.82&  3.26 & 6.50  & 2.38 & 1.89\\
\hline
\end{tabular}
\end{table}

Table \ref{tab2} gives the values of $a_s$, (energy
$\epsilon)/\sqrt{\kappa}$,\break (mass $M)/\sqrt{\kappa}$,
$\sqrt{\langle y^2\rangle}$, and $ \sqrt{\langle q_1^2\rangle
}/\sqrt{\kappa}$ of the lowest five states.  Fig.\ \ref{linV} displays
the energies as the horizontal lines, with their corresponding wave
functions $\Psi(y)$ exhibiting different number of nodes.  In the
two-body problem with the phenomenological two-body interaction, the
mass of the lowest state for the phenomenological potential is $M=
2.02 \sqrt{\kappa}$.  On the other hand, the Schwinger's exact
solution (\ref{2p1}) in field theory gives $M=e/\sqrt{\pi}$.
Therefore, by matching the mass of the lowest eigenstate from the
phenomenological two-body theory with the mass from Schwinger's field
theory, we find $\kappa$ to be
\begin{eqnarray}
\kappa=\frac{e^2}{4.08\pi}\sim \frac{e^2}{4\pi},
\label{4p9}
\end{eqnarray}
where for simplicity, we shall approximate the denominator 4.08$\pi$
to be $4\pi$.  We have thus obtained the phenomenological interaction
$\Phi$ in Eq.\ (\ref{4p1}) between two electric charges interacting in
QED in 1+1 dimensions.  Such a knowledge of the effective QED
interaction between two electric charges will enable us to study the
states of three quarks interacting in 1+1 dimensional QED.

We need the value of the coupling constant $e=e_\2d$ which can be
obtained from $e_\4d$.  In the physical world of 3+1 dimensions, the
one-dimensional open string without a structure is in fact an
idealization of a flux tube with a transverse radius $R_T$.  The
masses calculated in 1+1 dimensions can represent physical masses,
when the structure of the flux tube is properly taken into account.
Upon considering the structure of the flux tube in the physical 3+1
dimensions, we find that the coupling constant $e_{\rm 2D}$ in 1+1
dimensions is related to the physical coupling constants $e_{\4d}$ in
3+1 dimensions by the flux tube radius $R_T$,
\cite{Won10,Won20,Won09,Kos12,Kos21}
\begin{eqnarray}
(e_{\2d})^2=\frac{1}{\pi
    R_T^2}(e_{\4d})^2=\frac{4\alpha_{\4d}}{R_T^2},
\label{12}
\end{eqnarray}
with $\alpha_\4d=1/137$ and $R_T =$ 0.4 fm \cite{Won20}, which
yields $\sqrt{\kappa}=23.8$ MeV, and $\hbar/\sqrt{\kappa}=8.3$ fm.

As a final check of a faithful representation of the phenomenological
two-body model for the Schwinger exact field theory in 1+1 dimensions,
we note that they share the distinct property that the mass of the
system increases with the increase in the magnitude of the coupling
constant, in contrast to a non-confining interaction such as the
positronium where the mass of the composite system decreases with the
increase in the magnitude of the coupling constant.

We note from the above solutions of the two-body problem that a
fermion-antifermion composite system with the phenomenological QED
interaction possesses excited states with a higher number of nodes
$n$, in addition to the lowest state with $n=$ 0.  They represent
higher string vibrational excitations of the fermion-antifermion
system as an open string.
\begin{figure}[h]
\centering
\includegraphics[scale=0.38]{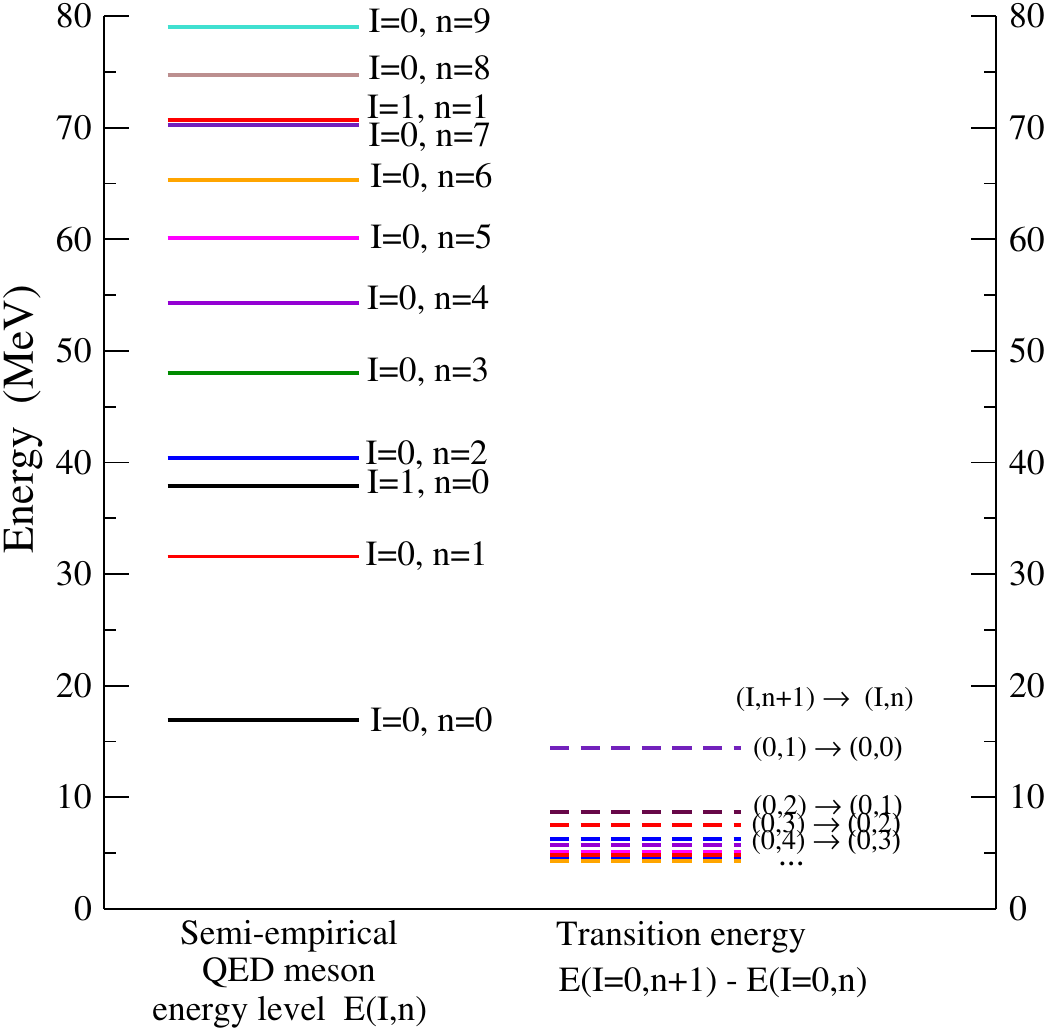}
\caption{ The left panel gives the semi-empirical QED meson energy
  level $E(I,n)$ where $I$ is the isospin quantum number and $n$ is the
  number of nodes, as obtained from Fig.\ \ref{linV} and Table I by
  assuming the observed X17 state at 16.70 MeV \cite{Kra16}, as the
  band heads of the $(I=1,n=0)$ and and the E38 state at $E=37.38$ MeV
  \cite{Abr19} as the band head of the $(I=1,I_3=0,n=0)$ states
  respectively.  The right panel gives the transition energy
  $E(I$=$0,n+1)-E(I$=$0,n)$ as dashed lines.  }
\label{spectrum}
\end{figure}
As a rough guide for future searches of QED meson states in $(q\bar
q)$ composite systems, we can treat the observed X17 state at $E=$
16.70 MeV \cite{Kra16} to be the lowest band heads of the $(I=0,n=0)$
state, and the E38 state at $E=$ 37.38 MeV \cite{Abr19} to be the
lowest band heads of the $(I=1,I_3=0,n=0)$ state respectively.  We can
then use the QED meson solutions as displayed in Fig.\ \ref{linV} and
Table \ref{tab2} to build on these two band heads semi-empirically an
approximate theoretical energy level diagram for QED meson states with
higher numbers of nodes $n$, as shown in Fig.\ \ref{spectrum}.  On the
right panel of Fig.\ \ref{spectrum}, we also display the decay energy
$E(I=0,n+1) - E(I=0,n)$ for the transition from the $(I=0,n+1)$ state to
the $(I=0,n)$ state.  This decay energy corresponds to the diphoton
energy if the $(I=0,n+1)$ state decays into the $(I,n)$ state with the
emission of two real or virtual photons as indicated in the diagram
Fig.\ \ref{deexcite}(b) and \ref{deexcite}(c) below.  Future
experimental searches for these two-body excited states and their
diphoton decays in QED mesons will be of great interest.  It should be
emphasized that the spectrum shown in Fig.\ \ref{spectrum} is useful
as an initial guide as many effects such as the fine structure of the
states and the deviations from the linear potential shape may modify
the spectrum.

\section { Variational Calculation for  the lowest two-body
  bound state  energy}

Before we apply the effective interaction $\Phi_{12}(x_{12})$ to the
three-quark problem, we wish to test here whether a variational
calculation using the effective interaction (\ref{4p1}) will also lead
to the same two-body bound state mass for the lowest-energy two-body
state.  The success of the variational calculations will pave the way
in a similar variational calculation for the three-body problem in
section 2.  We therefore evaluate the lowest two-body bound state
energy by using a Gaussian variational wave function.  For massless
quarks with $Q_1=1,$ $Q_2=-1$, we rewrite (\ref{4p4}) as
\begin{eqnarray}
&&\left\{{\cal H}_0 - E^2\right\} |\psi \rangle=0,  
\\
&&\text{where}~~~{\cal H}_0=- \frac{\partial^2}{\partial y^2}+ E|y|,~~~ {\rm and} ~~~E=\frac{\epsilon}{\sqrt{\kappa}}.
\label{4p15}
\end{eqnarray}
We introduce a Gaussian variational wave function with the variational
parameter $\sigma$,
\begin{eqnarray}
&&\Psi(y)\!=\!(\frac{1}{\sqrt{2\pi }\sigma})^{1/2} \exp\{-\frac{y^2}{4\sigma^2}\}\!=\!N\exp\{-\frac{y^2}{4\sigma^2}\}.
\end{eqnarray}
We obtain
\begin{eqnarray}
\langle {\cal H}_0\rangle (\sigma)=\frac{1}{4\sigma^2} +  \frac{2\sigma E}{\sqrt{2\pi }}.
\end{eqnarray}
From the requirement of 
$\delta \langle {\cal H}_0\rangle (\sigma)/\delta \sigma=0$,  we get 
\begin{eqnarray}
\sigma=\left (\frac{\sqrt{2\pi }}{4E}\right )^{1/3}\!\!\!,
\label{4p19}
\end{eqnarray}
at which
\begin{eqnarray}
 \langle {\cal H}_0\rangle=\frac{3 E}{\sqrt{2\pi }}.
\label{4.19}
\end{eqnarray}
From Eq.\ (\ref{4p15}), the value of $E^2$ at the equilibrium value of
$\sigma$ becomes
\begin{eqnarray}
E^2 =  \langle {\cal H}_0\rangle=\frac{3\sigma E}{\sqrt{2\pi }}.
\label{4p20}
\end{eqnarray}
Eliminating $\sigma$ from Eqs.\ (\ref{4p19}) and (\ref{4p20}) we get 
\begin{eqnarray}
E 
= \left (  \frac{3 }{(\sqrt{2\pi })^{2/3} 4^{1/3}}\right )^{3/4}
=   1.034
\label{Ap7}
\end{eqnarray}
\begin{eqnarray}
\sigma=\left (\frac{\sqrt{2\pi }}{4E}\right )^{1/3}=0.876.
\end{eqnarray}
The above variational calculation gives $E=\epsilon/\sqrt{\kappa}\sim1$ as
given in (\ref{Ap7}), and thus
\begin{eqnarray}
 \epsilon=\sqrt{\kappa},~~~\text{and}~~
M=2\epsilon=\frac{e}{\sqrt{\pi}},
\end{eqnarray}
which agrees with the lowest eigenenergy obtained by solving the wave
equation directly.  We find indeed that the variational calculation
can give the correct lowest-energy bound state mass.  This justifies
the use of the variational calculations in the three-body problem to
obtain the lowest-energy state of a QED neutron in section 2.

\section{The decay and detection of  composite  $q\bar q$ QED mesons}

\begin{figure}[H]
\centering
\includegraphics[scale=0.66]{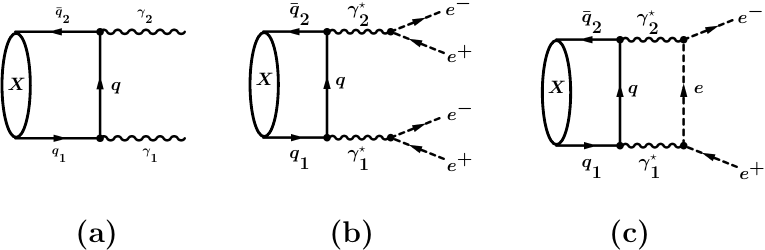}
\caption{ Fig.\ \ref{decay}(a) depicts the diagram for the decay of
  the QED meson $X$ into two real photons ($\gamma_1\gamma_2)$,
  Fig.\ \ref{decay}(b) the decay of the QED meson $X$ into two virtual
  photons ($\gamma_1^*\gamma_2^*)$, and Fig.\ \ref{decay}(c) the decay
  of the QED meson $X$ into a dilepton $(e^+e^-)$ pair.  }
\label{decay}
\end{figure}

We would like to review and extend here our knowledge on the decay of
the QED meson \cite{Won10,Won20} to facilitate the experimental
detection of QED mesons and the QED neutron.  We consider a $q\bar q$
composite system $X$ formed by a valence quark $q_1$ and a valence
antiquark $\bar q_2$ interacting with the effective QED interaction
$\Phi$.  As shown in Appendix D, there can be many eigenstate
solutions of the relativistic two-body equations for the composite
system $X$.  The additional intrinsic degrees of freedom of the quarks
will add further to the complexity of the spectrum and the number of
the composite particles $X$.  In the 1+1 dimensional description, the
composite particle $X$ cannot decay, as the quarks execute a yo-yo
motion along the string in an idealization of the flux tube, and the
photon is represented by an effective interaction binding the quarks.
In the physical 3+1 dimensions where the structure of the flux tube is
taken into account and the photon decay channel opens up, the quark
and the antiquark at different transverse coordinates in the tube
traveling from opposing longitudinal directions can make a sharp
change of their trajectories turning to the transverse direction to
annihilate, leading to the emission of photons as depicted in
Fig.\ \ref{decay}(a), \ref{decay}(b), and \ref{decay}(c).  The number
of emitted photons depends on the spin and parity of the decaying
system \cite{Lan48,Yan50}.  We have illustrated the case of two photon
decay in Fig.\ \ref{decay}(a), \ref{decay}(b), and \ref{decay}(c).
The emitted photon can be two real photons ($\gamma_1\gamma_2)$ in
\ref{decay}(a), two virtual photons ($\gamma_1^*\gamma_2^*)$, in
\ref{decay}(b), or a dilepton $(e^+e^-)$ pair in \ref{decay}(c).

\begin{figure}[h]
\centering
\includegraphics[scale=0.48]{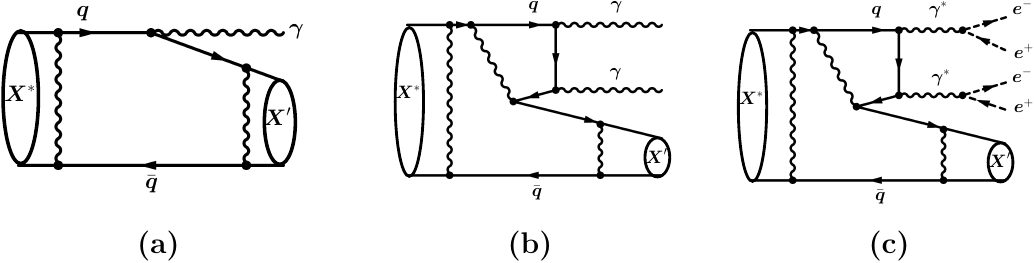}
\caption{ Diagrams for the de-excitation of the  excited QED meson $X^*$ to the QED meson $X'$ with particle emissions:  Fig.\ \ref{deexcite}(a)  $X^* \to X' + \gamma$ ,
 Fig.\ \ref{deexcite}(b)  $X^* \to X' + \gamma + \gamma$ , and 
 Fig.\ \ref{deexcite}(c)  $X^* \to X' + \gamma^*+ \gamma^*$ .
}
\label{deexcite}
\end{figure}

There can be excited QED meson states $X^*$ with nodal number $n$
which can de-excite to the lower QED meson state $X'$ with a smaller
nodal number $n'<n$, with the emission of $\gamma$, $\gamma\gamma$,
and $\gamma^* \gamma^*$ as shown in Fig.\ \ref{deexcite}.

\begin{figure}[H]
\centering 
\includegraphics[scale=0.40]{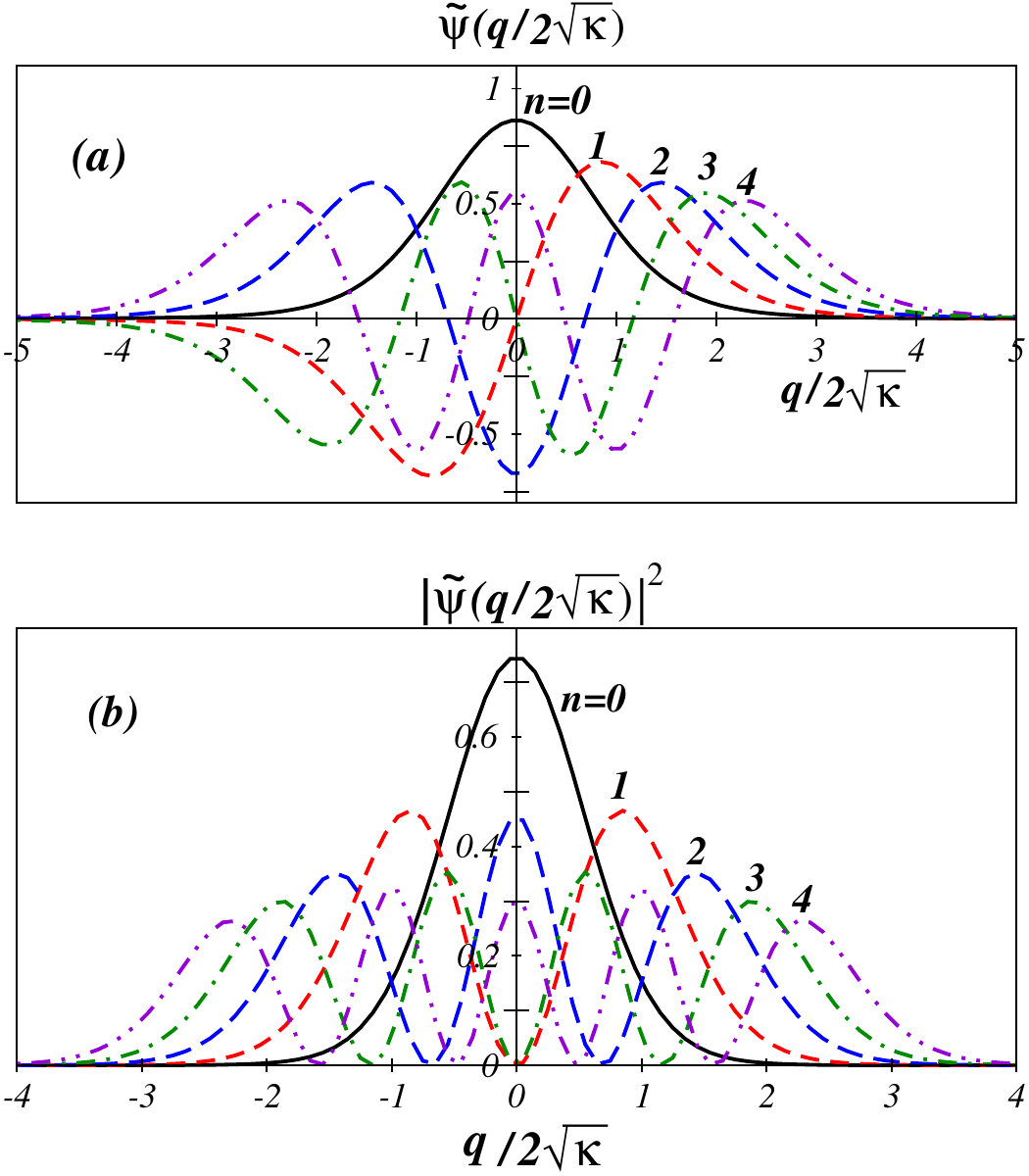}
\caption{Fig.\ \ref{figp}(a) gives the two-body wave function in
  momentum space $\tilde \Psi(q/2\sqrt{\kappa})$ of the composite
  $q$-$\bar q$ system as a function of the relative momentum
  $q=q_1-q_2=2q_1$ in units of $2\sqrt{\kappa}$, and
  Fig. \ref{figp}(b) gives $|\tilde \Psi(q/2\sqrt{\kappa})|^2$.  }
\label{figpp}
\end{figure}

For the decay from a bound state $X$, the decay amplitude needs to be
folded in with the bound state momentum wave function.  For $X\to
k_1+k_2$ where the final states $(k_1, k_2)$ are
$(\gamma_1,\gamma_2)$, $(\gamma_1^*,\gamma_2^*)$, or $(e^+,e^-)$ as in
diagrams\ \ref{decay}(b), \ref{decay}(c), and \ref{decay}(d), the
decay amplitudes are \cite{Cra06,Cra07}
\begin{eqnarray}
M(X\!\!\to \!k_1 + k_2) \! =\!\!\! \int \!\!d^3\! q_1 \tilde \Psi ( q_1)
\Gamma (q_1\!+\!\bar q_2 \!\to\! k_1 \!+\! k_2),
\end{eqnarray}
where $\tilde \Psi (q_1)$ is the bound state momentum wave function of
a constituent, $q_1=q_{p_{q_1}p_{\bar q_2}}/2$, and $\Gamma (q_1+\bar
q_2 \to k_1 + k_2)$ is the Feynman amplitude for the diagram of
$q_1+\bar q_2 \to k_1 + k_2$, (see e.g. Eqs.\ (3.6) of \cite{Cra06} or
Eq.\ (2.2) of \cite{Cra07} for the case of two-photon decay).
The two-body spatial wave function $\Psi(y)$ for the QED meson $X$ in
Eqs.\ (\ref{4p4}) has been expressed in terms of the dimensionless
relative spatial coordinate $y=\sqrt{\kappa}(x_{q_1}-x_{\bar q_2})$.
The corresponding wave function in the momentum space, $\tilde
\Psi(q/2\sqrt{\kappa})$ is the Fourier transform of $\Psi(y)$ of
Eqs.\ (\ref{4p3}) and (\ref{4p6}),
\begin{eqnarray}
\tilde \Psi(q/2\sqrt{\kappa})\!=\!\tilde \Psi(q_1/\sqrt{\kappa})
\!=\!\frac{1}{\sqrt{2\pi} } \!\!\int\!\! e^{-i (q_1/\sqrt{\kappa}) y }
\Psi(y) dy ,
\end{eqnarray}
where $q=q_1-q_2=2q_1$, and the constituent momentum $q_1$ is expressed
in units of $\sqrt{\kappa}$.  
The momentum distribution of the
composite $q_1\bar q_2$ system $X$ with a mass $M$ is
\begin{eqnarray}
\frac{dN(P,q)}{dP dq} = \delta (P^2-M^2) |\tilde
\Psi(q/2\sqrt{\kappa})|^2.
\label{5p3}
\end{eqnarray}
We show in Figs.\ \ref{figpp}(a) the momentum wave functions $\tilde
\Psi(q/2\sqrt{\kappa})$, and in Fig.\ \ref{figpp}(b) the corresponding
momentum probability function $|\tilde \Psi(q/2\sqrt{\kappa})|^2$, for
the lowest five states of the composite system.  We can understand the
contents of Eq.\ (\ref{5p3}) intuitively as follows: the invariant
square of the momentum sum, $P^2$, probes the invariant mass $M^2$ of
the composite QED meson, whereas the invariant square of the momentum
difference, $q^2$, measures the inverse spatial size of the decaying
QED meson as $ \hbar/ \sqrt{\langle q^2\rangle }$, and the number of
nodes of $|\tilde \Psi(q/2\sqrt{\kappa})|^2$ in Eq.\ (\ref{5p3}) and
Fig.\ \ref{figpp}(b) reveals the internal structure of the composite
$q_1{\bar q}_2$ system.

The decay via two virtual photons  in diagram Fig.\ \ref{decay}(b)
may provide an interesting probe to yield additional information on
the decaying QED meson parent particle.  One can construct the
invariant momenta square of the sum and differences of the 4-momenta
of the two virtual photons,
\begin{eqnarray}
P_{\gamma_1^* \gamma_1^*}^2&&=(p_{\gamma_1^*}+p_{\gamma_2^*})^2,
\nonumber\\
Q_{\gamma_1^* \gamma_1^*}^2&&=- (p_{\gamma_1^*}-p_{\gamma_2^*})^2,
\end{eqnarray}
and $P\equiv P_{\gamma_1^* \gamma_1^*}=\sqrt{P_{\gamma_1^* \gamma_2^*}^2}$, 
$Q \equiv Q_{\gamma_1^* \gamma_1^*}=\sqrt{-q_{\gamma_1^* \gamma_2^*}^2}$.   Experimental measurement of the virtual diphoton pair distribution 
\begin{eqnarray}
\frac{dN(P_{\gamma_1^* \gamma_2^*} ,q_{\gamma_1^* \gamma_2^*}^* )}{dP_{\gamma_1^* \gamma_2^*} dQ_{\gamma_1^* \gamma_2^*} } =\frac{dN(P,Q)}{dP~dQ} 
\label{5p4x}
\end{eqnarray}
will provides useful information on the composite $q\bar q$ particles
of massless light quarks from virtual diphoton decay measurements.
Specifically, one makes a scatter plot of diphoton events on the two
dimensional plane of $P$ and $Q$.  In actual experiments with the
presence of cuts and windows in various kinematical regions, one may
resort to the use of the event mixing method to normalize the
distribution as
\begin{eqnarray}
\frac{dN(P,Q)}{dP ~dQ}\!\biggr |_{\rm norm}\hspace{-0.2cm}
~~\equiv \frac
{  \left [\frac{dN(P,Q)}{dP~ dQ} \right ]_{\rm correlated}}
{~~~\left [ \frac{dN(P,Q )}{dP~dQ }\right ]_{\rm mixed ~events}}.
\end{eqnarray}
The above normalized distribution ${dN(P,Q )/}{dP\,dQ} |_{\rm norm}$
is expected to cluster sharply around the invariant mass $M=(P+Q)/2$
of the composite particles and spread out in width in the direction of
$Q$.  One can alternative display the distribution in terms of the
rotated coordinates $(P+Q)/2$ and a small stripe of $P-Q >0$.

\end{appendix}

\end{document}